\documentstyle[12pt,epsf]{article}

\def\fnum@figure{Fig. \thefigure}
\def\fnum@table{Tabella \thetable}

\def\thebibliography#1{\section*{
  {References}}\list
{[\arabic{enumi}]}{\settowidth\labelwidth{[#1]}\leftmargin\labelwidth
    \advance\leftmargin\labelsep
    \usecounter{enumi}}
    \def\newblock{\hskip .11em plus .33em minus -.07em}
    \sloppy
    \sfcode`\.=1000\relax}

\font\Bbbfont=msbm10
\def\Bbb#1{\hbox{\Bbbfont#1}}

\begin{document}

\begin{center}
{ \Large \bf  Hierarchies of invariant spin models}

\vspace{24pt}
\end{center}

\begin{center}
{\large
{\sl Gaspare Carbone}$\,^{a,}$\footnote{email carbone@sissa.it},
{\sl Mauro Carfora}$\,^{b,}$\footnote{email Mauro.Carfora@pv.infn.it},
and
{\sl Annalisa Marzuoli}$\,^{b,}$\footnote{email Annalisa.Marzuoli@pv.infn.it}
}
\vspace{24pt}

$^a$~S.I.S.S.A.-I.S.A.S.,\\
Via Beirut 2-4, 34013 Trieste, Italy

\vspace{12pt}
$^b$~Dipartimento di Fisica Nucleare e Teorica, \\
Universit\`a degli Studi di Pavia, \\
via A. Bassi 6, I-27100 Pavia, Italy, \\
and\\
Istituto Nazionale di Fisica Nucleare, Sezione di Pavia, \\
via A. Bassi 6, I-27100 Pavia, Italy

\end{center}

\vspace{12pt}

\begin{center}
{\bf Abstract}
\end{center}

\vspace{12pt}

In this paper we present classes of state sum models based on the recoupling 
theory 
of angular momenta of $SU(2)$ (and of its 
$q$-counterpart $U_q(sl(2))$, $q$ a root 
of unity).
Such classes are arranged in hierarchies depending on the dimension $d$, and 
include 
all known  closed models, {\em i.e.} the Ponzano--Regge state sum and the 
Turaev--Viro invariant in dimension $d=3$, the Crane--Yetter invariant in 
$d=4$. In general, the recoupling coefficient associated with a $d$-simplex 
turns out to be a $\{3(d-2)(d+1)/2\}j$ symbol, or its $q$-analog.\par
\noindent Each of the state sums can be further extended to compact 
triangulations
$(T^d,\partial T^d)$ of a $PL$-pair 
$(M^d,\partial M^d)$, where the triangulation
of the boundary manifold is not keeped fixed. In both cases we find out the
algebraic identities which translate complete sets of topological moves, thus 
showing 
that all state sums are actually independent of the particular triangulation 
chosen. 
Then, owing to Pachner's theorems, it turns out that classes of $PL$-invariant 
models can be defined in any dimension $d$.\par

\begin{flushleft}

\end{flushleft}

\vfill

\newpage

\section{Introduction}

In what follows we shall always consider either
{\em closed}  $d$-dimensional simplicial $PL$-manifolds or {\em compact} 
$d$-dimensional simplicial $PL$-pairs $(M^d, \partial M^d)$, where the 
triangulation on the boundary
$(d-1)$-manifold is not keeped fixed.\par
\noindent Recall that a closed $PL$-manifold of dimension $d$ is a polyhedron 
$M^d \cong |T^d|$, each point of which has a neighborhood, in $M^d$,
$PL$-homeomorphic to an open set in ${\Bbb R}^d$.
The symbol $\cong$ denotes homeomorphism, $T^d$ is the underlying (finite) 
simplicial complex and $|T^d|$ denotes the associated 
topological space. $PL$-manifolds are realized by simplicial manifolds under the 
equivalence relation generated by 
$PL$-homeomorphisms. In particular,
two $d$--dimensional closed $PL$-manifolds $M_1^d \cong |T_1^d|$ and $M_2^d \cong 
|T_2^d|$ are $PL$-homeomorphic, or
$M_1^d \:\cong_{PL} \:M_2^d$, 
if there exists a map $g:M_1^d \rightarrow M_2^d$ which is both a homeomorphism 
and a simplicial
isomorphism. We shall use the notation\par

\begin{equation}
T^d \:\longrightarrow \: M^d\:\cong \:|T^d|
\label{clotri}
\end{equation}

\noindent to denote a {\em particular triangulation} of the closed $d$-dimensional 
$PL$-manifold $M^d$.\par
\noindent A $PL$-invariant $Z[M^d]$ is a quantity which is independent of the 
particular 
triangulation chosen in (\ref{clotri}). The value of such an invariant depends 
just on
the $PL$-class of the closed manifold , namely it is the same for 
$PL$-homeomorphic
manifolds. The previous definitions can be naturally extended
to the case of a  $PL$-pair $(M^d,\partial M^d)$ of dimension $d$ according to:

\begin{equation}
(T^d, \partial T^d )\:\longrightarrow \: (M^d, \partial M^d)\:\cong \:(|T^d|, 
|\partial T^d|),
\label{bountri}
\end{equation}

\noindent where $\partial T^d$ is the unique triangulation 
induced on the $(d-1)$-dimensional boundary $PL$-manifold $\partial M^d$ by the 
chosen
triangulation $T^d$ in $M^d$. A $PL$-invariant $Z[(M^d, \partial M^d)]$ is a 
quantity which is independent of the particular triangulation chosen in 
(\ref{bountri}). 
The value of such an invariant depends upon
the $PL$-class of the pair, namely it is the same for $PL$-homeomorphic pairs 
(the reader may refer {\em e.g.} to \cite{rourke} for more details on 
$PL$--topology).\par
\vskip 0.5 cm
The general setting of the content of this paper can be summarized as follows:\par
\begin{itemize}
\item {\em step 1}) Given a (suitable defined) $PL$-invariant state sum 
$Z[M^{d-1}]$ 
for a closed $(d-1)$-dimensional $PL$-manifold $M^{d-1}$, we extend it to a state 
sum for 
a pair $(T^d, \partial T^d \equiv T^{d-1})$. This is achieved by assembling in a 
suitable way 
the {\em squared roots} of the symbols associated with the fundamental blocks in
$Z[M^{d-1}]$ in order to pick up the recoupling symbol to be  associated with the 
$d$-dimensional simplex; the dimension of the $SU(2)$-labelled (or the 
$q$-colored)   
$(d-2)$-simplices is keeped fixed when passing from $T^{d-1}$ to $T^d \supset 
T^{d-1}$.
\item {\em step 2}) The state sum for $(T^d, \partial T^d)$ gives rise to a 
$PL$-invariant
$Z[(M^d, \partial M^d)]$ owing to the fact that we can exploit a set of 
topological moves,
the {\em elementary shellings} of Pachner \cite{pachner90} (the algebraic 
identities 
associated with such moves in $d=3$ were established 
in \cite{carbone} and \cite{convegno}).
\item {\em step 3}) From the expression of $Z[(M^d, \partial M^d)]$ we can now 
extract
a state sum for a {\em closed} triangulation $T^d$. The proof of its 
$PL$-invariance 
relies now on the algebrization in any dimension $d$ of the {\em bistellar moves} 
introduced in \cite{pachner87}. The procedure turns out to be consistent with 
known 
results in dimension $3$ (see \cite{ponzano}, \cite{turaev} and \cite{carter95}) 
and
in dimension $4$ (see \cite{ooguri4d} and \cite{crane}), and provides us with a 
$PL$-invariant $Z[M^d]$ where each $d$-simplex in $T^d$ is represented by a 
$\{3(d-2)(d+1)/2\}j$ 
recoupling coefficient of $SU(2)$ (or by the corresponding $q$-analog).   
\end{itemize}

The scheme we have outlined above gives us an algorithmic procedure for
generating (different kinds of) invariants for closed manifolds in contiguous 
dimensions, 
namely $Z[M^{d-1}] \rightarrow Z[M^d]$. 
Futhermore, the (multi)--hierarchic structure underlying these
classes of invariants is sketched below as an array:\par

\vskip 0.5 cm
\begin{center}
$\begin{array}{cccccccc}
dimension: & 2 & 3 & 4 & \ldots & d & d+1 \\
  &  &  &  &  &  &   &  \\
  & Z^2_{\chi} & Z^3_{PR} &  &  &  &  \\
  &  &  &  &  &  &   &  \\
  &  & Z^3_{\chi} & Z^4_{CY} &  &  &  \\
  &  &  & Z^4_{\chi} & \ddots &  &  \\
  &  &  &  & \ddots & \ddots &  \\
  &  &  &  &  &  Z^d_{\chi} & Z^{d+1}  \\
  &  &  &  &  &  &   &  \\
  &  &  &  &  &  & Z^{d+1}_{\chi}
\end{array}$
\end{center}
\vskip 0.5 cm
\noindent The quantities $Z^d_{\chi} \equiv Z_{\chi}[M^d]$ on the first diagonal
of the array, which we referred to in {\em step 1)}, are invariants depending 
upon the Euler characteristic of the 
closed manifold $M^d$; they can be defined both in the classical case of $SU(2)$
and in the case $q \neq 1$,(in this last case the notation
$Z^d_{\chi}(q)$ should be more suitable). They are
obtained in any dimension $d$ by labelling the $(d-1)$-simplices 
of the triangulations with the ranks of $SU(2)$ representations, 
$j=0,1/2,1,3/2,\ldots$ (see {\em Section 4} 
for the general definition in the $q$-case). Notice that these invariants 
are in fact trivial in dimension $d=2n+1$ since here we are dealing exclusively 
with manifolds.\par
\noindent The hierarchy on the second diagonal of the array includes classical
$PL$-invariants $Z^d \equiv Z[M^d]$ involving products of $\{3(d-2)(d+1)/2\}j$ 
symbols 
of $SU(2)$ as we said in {\em step 3)}. In this case the labelling $j$ have to be 
assigned 
to the $(d-2)$-simplices
of each triangulation (namely edges in $d=3$, triangles in $d=4$, and so on). 
Thus we recover the Ponzano--Regge model $Z^3_{PR}$ and the Ooguri--Crane--Yetter 
invariant
$Z^4_{CY}$($q=1$); the other invariants are new.
A similar remark holds true also in the $q$-deformed context, where the hierarchy
would be rewritten in terms of the counterparts $Z^d(q)\equiv Z[M^d](q)$ (and in 
particular we found the Turaev--Viro invariant in $d=4$, see \cite{turaev}).\par
\noindent Coming back to the relations between invariants lying in the same row
of the array, they can be further analyzed in view of the extension of each
$Z[M^d]$ to $Z[(M^d, \partial M^d)]$. Thus the first row can be read as a
$PR$-model with $Z^2_{\chi}$ on its boundary, the second row as a      
$CY$-model with $Z^3_{\chi}$ on its boundary, while the other rows display
new invariants for $PL$-pairs in each dimension.\par
\noindent As a consequence of the above remarks, the whole table (together with
a similar $q$-table) can be reconstructed row by row just from the explicit form
of the invariant $Z^d_{\chi}$. Then, in a sense, it is not surprising that the 
invariant $Z^4_{CY}$, having on its boundary $Z^3_{\chi} = const.$ for any choice 
of $\partial M^4$,  turns out to be simply the discretized version of a 
combination of signature and Euler characteristic of $M^4$ (see \cite{roberts}). 
On the other hand, the invariants 
$Z^{2n+1}$ generated by non--trivial $Z^{2n}_{\chi}$ are expected to be related to 
suitable types of torsions, as happens in the $3$-dimensional case (see {\em e.g.}
\cite{turlibro}). However, the proper way to investigate the nature of the 
invariants in $d=2n > 4$ and $d=2n+1 > 3$ is by no doubts the search for
explicit correspondences with some $TQFT$s. It turns out that the continuous 
counterparts of the
classical $Z^d$ are indeed $BF$ theories, although in the even case the 
identification of the
resulting invariant(s) does not seem so straightforward. Similar considerations 
apply to the   
$Z^d(q)$, which should be discretized versions of $BF$ theories with suitable
cosmological terms. These last issues will be discussed elsewhere.\par
\vskip 0.5 cm
For the sake of clarity in the exposition, the presentation does not follow 
exactly
the schedule given in {\em steps 1--3} at the beginning, mainly
owing to the fact that calculations in dimension $d > 4$ can be performed only by
diagrammatical methods. Thus, as an illustration of the analytical approach,
we present in {\em Section 2} a short rewiew of \cite{carbone} and 
\cite{convegno},
while in {\em Section 3} we provide the extension of the $CY$-invariant ($q=1$)
to the case of a pair $(M^4, \partial M^4)$ and the expression of the induced 
$Z^3_{\chi}$ on the boundary. In the following {\em Section 4} we give the 
explicit 
form of $Z^d_{\chi}$ for any closed $M^d$. In {\em Section 5} we generate 
$Z[(M^d, \partial M^d)]$ from $Z^{d-1}_{\chi}$ and we show its $PL$-invariance. 
{\em Section 6} contains 
the proof that the state sum induced by $Z[(M^d, \partial M^d)]$ when 
$\partial M^d = \emptyset$ is a well defined $PL$-invariant. Finally, in 
{\em Appendix B} we collect the basic notations in view of the extension 
to the $Z^d(q)$ hierarchy.\par 
\vfill

\newpage


\section{Ponzano--Regge model for $(M^3, \partial M^3)$ and \\
induced $2$-dimensional invariant}
Following \cite{carbone} and \cite{convegno}, the connection between a recoupling 
scheme 
of $SU(2)$ angular momenta and the combinatorial structure of a compact, 
$3$-dimensional simplicial pair $(M^3, \partial M^3)$ can be established by 
considering 
{\em colored} triangulations which
allow us to specialize the map (\ref{bountri}) according to\par
\begin{equation}
(T^3(j), \partial T^3 (j', m))\:\longrightarrow\:(M^3, \partial M^3).
\label{coltri}
\end{equation}

\noindent This map represents a triangulation associated with an admissible 
assignment of both spin variables to the collection of the edges 
($(d-2)$-simplices) 
in $(T^3, \partial T^3)$ and of momentum projections to the subset of edges lying 
in 
$\partial T^3$. 
The collective variable $j \equiv \{j_A\}$, $A=1,2,\ldots,N_1$, denotes all the 
spin variables, $n'_1$ of which are associated
with the edges in the boundary (for each $A$: $j_A=0,1/2,1,3/2,\ldots$ in  $\hbar$ 
units).
Notice that the last subset is labelled both by $j' \equiv \{j'_C \}$, 
$C=1,2,\ldots,n'_1$,
and by $m \equiv \{m_C\}$, where $m_C$ is the projection of $j'_C$ along the fixed 
reference axis (of course, for each $m$,
$-j \leq m \leq j$ in integer steps). The consistency in the assignment of the 
$j$, $j'$, $m$ variables is ensured if we require 
that\par 
\begin{itemize}
\item each $3$-simplex $\sigma_B^3$, ($B=1,2,\ldots,N_3$), 
in $(T^3, \partial T^3)$ must be associated, apart from a phase factor, with a 
$6j$ symbol of $SU(2)$, namely\par
\begin{equation}
\sigma^3_B \:\longleftrightarrow \:(-1)^{\sum_{p=1}^6 j_p} \:
\left\{ \begin{array}{ccc}
j_1 & j_2 & j_3 \\
j_4 & j_5 & j_6
\end{array}\right\}_B;
\label{jtresim}
\end{equation}
\item each $2$-simplex
$\sigma_D^2$, $D=1,2,\ldots,n'_2$ in $\partial T^3$ must be associated with a 
Wigner $3jm$ symbol of $SU(2)$ according to\par
\begin{equation}
\sigma_D^2 \:\longleftrightarrow\: (-1)^{(\sum_{s=1}^3 m_s)/2}
\left( \begin{array}{ccc}
j'_1 & j'_2 & j'_3 \\
m_1 & m_2 & -m_3
\end{array}\right)_D.
\label{jmduesim}
\end{equation}
\end{itemize}
\noindent Then the following state sum can be defined\par
\begin{eqnarray}
\lefteqn{ Z^3_{PR} \equiv Z_{PR}[(M^3, \partial M^3)]\,=}\nonumber\\
& & =\,\lim_{L\rightarrow \infty}\:
\sum_
{\left \{\begin{array}{c}
(T^3, \partial T^3)\\ 
j,j',m \leq L
\end{array}\right\}}
Z[(T^3(j),\partial T^3(j',m)) \rightarrow (M^3, \partial M^3); L],
\label{clsum1}
\end{eqnarray} 
\noindent where\par
\begin{eqnarray}
\lefteqn{Z[(T^3(j),\partial T^3(j',m)) \rightarrow (M^3, \partial M^3); L] 
=}\hspace{.5in}\nonumber \\
& & =\Lambda(L)^{-N_0}\,\prod_{A=1}^{N_1} (-1)^{2j_A} (2j_A+1)\,\prod_{B=1}^{N_3} 
(-1)^{\sum_{p=1}^6 j_p}
\left\{ \begin{array}{ccc}
j_1 & j_2 & j_3 \\
j_4 & j_5 & j_6
\end{array}\right\}_B \cdot  \nonumber \\
& & \cdot \,\prod_{D=1}^{n'_2} (-1)^{(\sum_{s=1}^3 m_s)/2}
\left( \begin{array}{ccc}
j'_1 & j'_2 & j'_3 \\
m_1 & m_2 & -m_3
\end{array}\right)_D. 
\label{clsum2}
\end{eqnarray}
\noindent $N_0$, $N_1$, $N_3$ denote respectively the total number of vertices, 
edges and tetrahedra in $(T^3(j),
\partial T^3(j',m))$, while $n'_2$ is the number of $2$-simplices lying in 
$\partial T^3(j',m)$. Notice that there appears 
a factor $\Lambda(L)^{-1}$ for each vertex in $\partial T^3(j',m)$, with $\Lambda 
(L)\equiv 4L^3/3C$, $C$ an arbitrary constant.\par
\noindent 
The state sum given in (\ref{clsum1}) and (\ref{clsum2}) 
when $\partial M^3 =\emptyset$ reduces to the usual Ponzano--Regge partition 
function for a 
closed manifold $M^3$; in such a case, it can be rewritten simply as\par
\begin{equation}
Z_{PR}[M^3]\,=\,\lim_{L\rightarrow \infty}\:\sum_{\{T^3(j), j \leq 
L\}}\:Z[T^3(j)\rightarrow M^3; L],
\label{prsum1}
\end{equation}
\noindent where the sum is extended to all assignments of spin variables such that 
each of them is not greater than
the cut--off $L$, and each term under the sum is given by\par
\begin{equation}
Z[T^3(j)\rightarrow M^3; L] =\Lambda(L)^{-N_0}\,\prod_{A=1}^{N_1}(-1)^{2j_A} 
(2j_A+1)\,\prod_{B=1}^{N_3} (-1)^{\sum_{p=1}^6 j_p}
\left\{ \begin{array}{ccc}
j_1 & j_2 & j_3 \\
j_4 & j_5 & j_6
\end{array}\right\}_B.
\label{prsum2}
\end{equation}
\noindent As is well known, the above state sum gives the semiclassical partition 
function of Euclidean $3$-gravity with an action discretized according to Regge's 
prescription \cite{regge}. Moreover, it is formally invariant under any finite set 
of topological transformations performed on $3$-simplices in $T^3(j)$: following 
Pachner \cite{pachner87}, 
they are commonly known as {\em bistellar moves}.
It is a classical result  (see {\it e.g.} \cite{ponzano} and \cite{carter95}) 
that such moves can be expressed algebraically in terms of the Biedenharn--Elliott 
identity
(representing the moves ($2$ tetrahedra) $\leftrightarrow$ ($3$ tetrahedra))
and of both the B-E identity and the 
orthogonality conditions for $6j$ symbols, which represent the barycentic move 
together with its inverse, namely ($1$ tetrahedron) $\leftrightarrow$ ($4$ 
tetrahedra) (see \cite{russi} for the explicit expressions of these identities as 
well as for notations concerning other (re)coupling coefficients).\par
\noindent In general, if we denote by $n_d$ the number of $d$-simplices $\in T^d$ 
involved in a given bistellar operation, then such a move can be represented with 
the notation\par
\begin{equation}
[n_d \rightarrow (d+1)-(n_d-1)]^{{\bf d}}_{bst}
\label{dbst}
\end{equation} 
\noindent and the entire set of allowed moves in dimension $d$ is found for $n_d = 
1,2,\ldots,d+1$ (as an example, the barycentric subdivision corresponds to the 
case
$n_d=1$).\par
\noindent The invariance under bistellar moves is related to the $PL$-equivalence 
class of the manifolds involved. Indeed, Pachner proved in \cite{pachner87} that 
two closed $d$-dimensional $PL$-manifolds are $PL$-homeomorphic
if, and only if, their underlying triangulations are related to each other by a 
finite sequence of bistellar moves. Thus
in particular the state sum (\ref{prsum1}) is formally an invariant of the  
$PL$-structure of $M^3$ (the regularized counterpart being the Turaev--Viro 
invariant found in \cite{turaev}).\par
\noindent Turning now to the non trivial case of (\ref{clsum1}) with $\partial M^3 
\neq  \emptyset$, new types of topological transformations have to be taken into 
account. 
Indeed Pachner introduced moves which are suitable in the case of compact 
$d$-dimensional $PL$-manifolds with a non--empty boundary, the {\em elementary 
shellings} (see 
\cite{pachner90}). This kind of operation 
involves the cancellation of one $d$-simplex at a time in a given triangulation 
$(T^d, \partial T^d) \rightarrow
(M^d, \partial M^d)$ of a compact $PL$-pair of dimension $d$. In order to be 
deleted, the $d$-simplex must have some of its $(d-1)$-dimensional faces
lying in the boundary $\partial T^d$.  Moreover, for each elementary shelling 
there exists an inverse move which corresponds to the attachment of a new 
$d$-simplex to a suitable 
component in $\partial T^d$. It is possible to classify  the two sets of moves 
by setting\par
\begin{eqnarray}
[n_{d-1} \rightarrow d-(n_{d-1}-1)]^{{\bf d}}_{sh}\:\:\:\:,\:\:\:\:
[n_{d-1} \rightarrow d-(n_{d-1}-1)]^{{\bf d}}_{ish}\:\:,
\label{dshel}
\end{eqnarray}
\noindent where $n_{d-1}$ represents the number of $(d-1)$-simplices 
(belonging to a single $d$-simplex)
involved in an elementary shelling and in an inverse shelling, respectively. 
Then the full set of operations is found when $n_{d-1}$ runs over
$(1,2,\ldots,d)$ in both cases.\par
\noindent In \cite{carbone} identities representing the three types 
of elementary shellings (and their inverse moves)
for the $3$-dimensional triangulation given in (\ref{coltri}) were 
established.\par
\noindent The first identity represents, according to (\ref{dshel}), the move 
$[2 \rightarrow 2]^{{\bf 3}}_{sh}$. The topological content of this identity 
is drawn on the top of FIG.1, 
\begin{figure}[htb]
\leavevmode
\hspace{1.5cm} \epsfbox{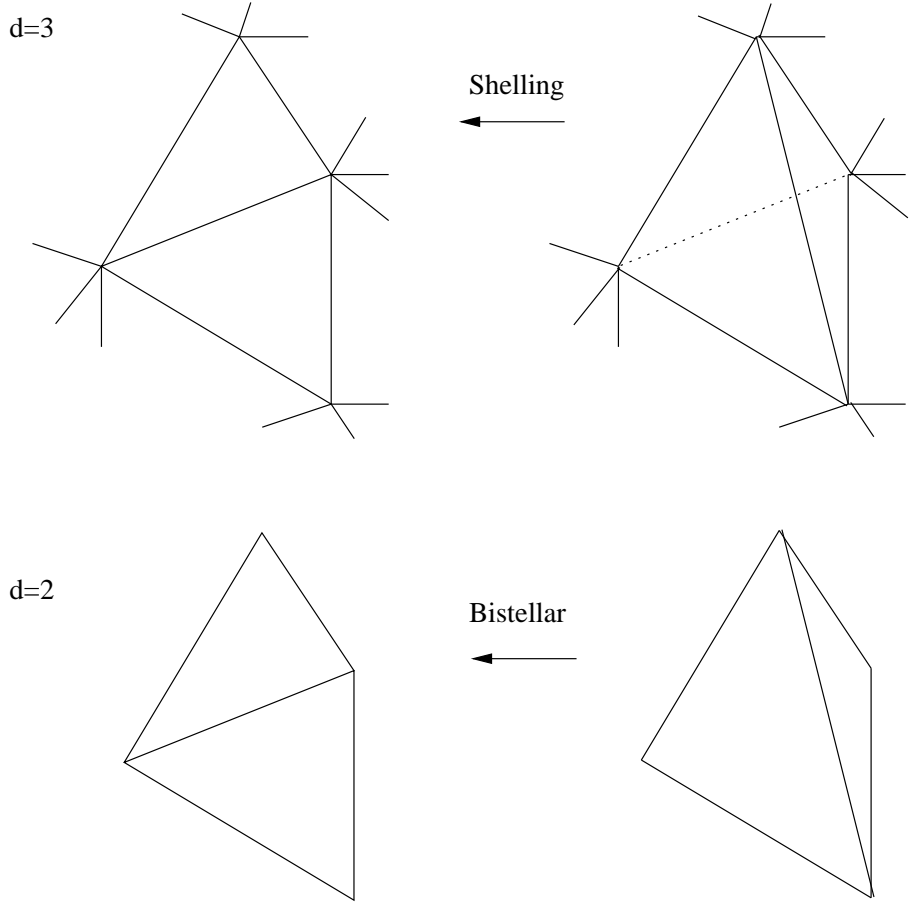}
\caption{The shelling $[2 \rightarrow 2]^{{\bf 3}}_{sh}$ and the corresponding 
bistellar $[2 \rightarrow 2]^{{\bf 2}}_{bst}$.}
\label{shell3}
\end{figure}
while its formal expression reads\par 
\begin{eqnarray}
\lefteqn{\sum_{c\gamma} (2c+1)(-1)^{2c-\gamma}
\left( \begin{array}{ccc}
a & b & c \\
\alpha & \beta & \gamma
\end{array} \right)
\left( \begin{array}{ccc}
c & r & p \\
-\gamma & \rho ' & \psi
\end{array} \right)
 (-1)^{\Phi}\,
\left\{ \begin{array}{ccc}
a & b & c \\
r & p & q
\end{array}\right\}\,=}\hspace{1.5cm}\nonumber\\
& & =\,(-1)^{-2\rho}
\sum_{\kappa}\,(-1)^{-\kappa}
\left( \begin{array}{ccc}
p & a & q \\
\psi & \alpha & -\kappa
\end{array} \right)
\left( \begin{array}{ccc}
q & b & r \\
\kappa & \beta & -\rho '
\end{array} \right),
\label{clrodue}
\end{eqnarray}
\noindent where Latin letters $a,b,c,r,p,q,\ldots$ denote
angular momentum variables, Greek letters $\alpha, \beta, \gamma, \rho, 
\psi, \kappa,\ldots$ are the corresponding momentum projections 
and $\Phi \equiv a+b+c+r+p+q$.\par
\noindent Notice that in this section we agree that all $j$ variables appearing 
in $3jm$ symbols are associated with edges lying
in $\partial T^3$ in a given configuration,
while $j$ arguments of the $6j$ may belong either to $\partial T^3$ (if they 
have a counterpart in the nearby $3jm$) or to $int(T^3)$.\par
\noindent The other identities can be actually derived
(up to suitable regularization factors) from (\ref{clrodue}) 
and from both the orthogonality conditions for the $6j$ symbols and
the completeness conditions for the $3jm$ symbols (see \cite{russi}).
In particular, the shelling $[1 \rightarrow 3]^{{\bf 3}}_{sh}$ is sketched 
on the top of FIG.2 
\begin{figure}[htb]
\leavevmode
\hspace{4.5cm} \epsfbox{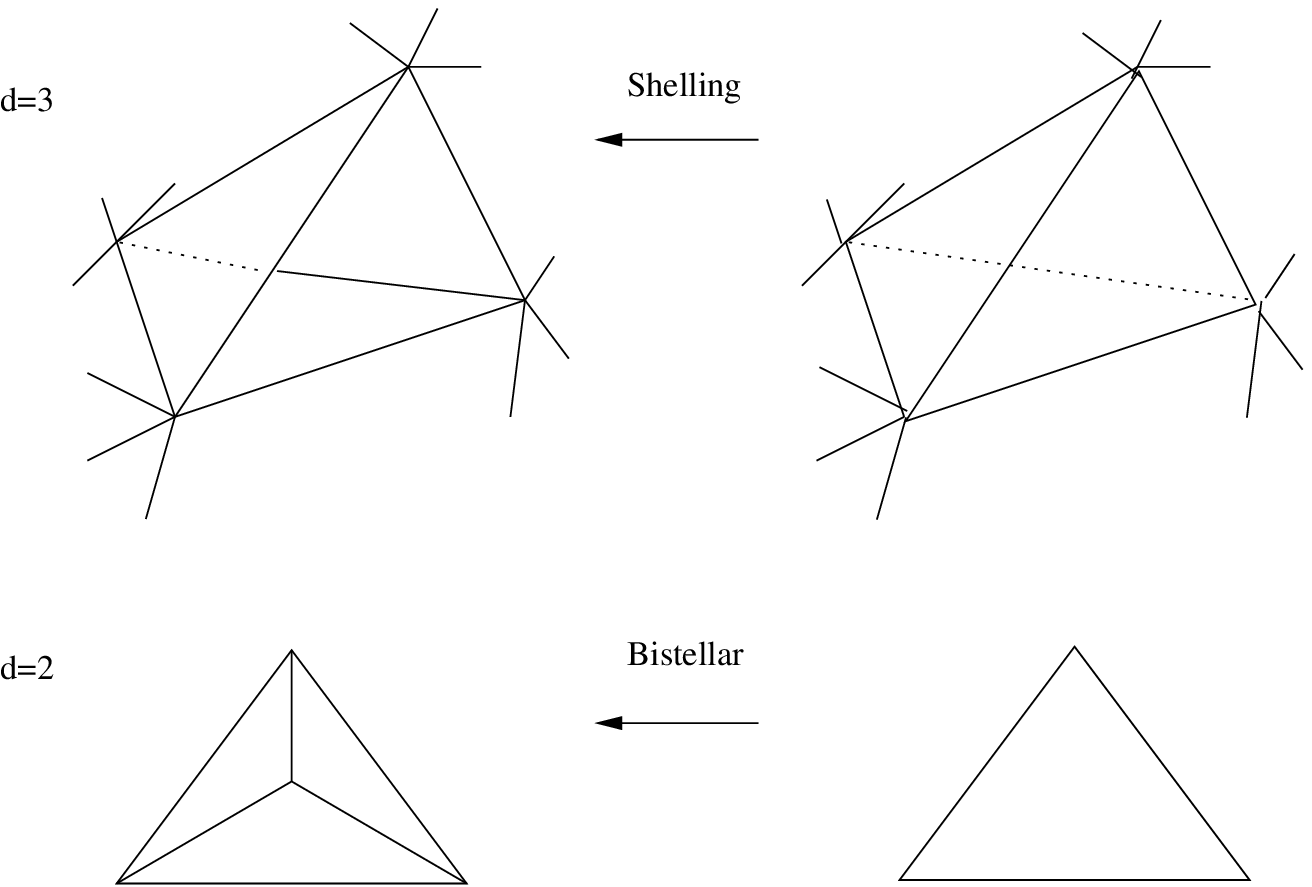}
\caption{ The shelling $[1 \rightarrow 3]^{{\bf 3}}_{sh}$ and the corresponding 
bistellar $[1 \rightarrow 3]^{{\bf 2}}_{bst}$.}
\label{shell12}
\end{figure}
and the corresponding identity is given by\par
\begin{eqnarray}
\lefteqn{\left( \begin{array}{ccc}
a & b & c \\
\alpha & \beta & \gamma
\end{array} \right)
(-1)^{\Phi}\,
\left\{ \begin{array}{ccc}
a & b & c \\
r & p & q
\end{array}\right\}\,=}\hspace{.5cm}\nonumber\\
& & =\sum_{\kappa \psi \rho}\,(-1)^{-\psi -\kappa -\rho}
\left( \begin{array}{ccc}
p & a & q \\
\psi & \alpha & -\kappa
\end{array} \right)
\left( \begin{array}{ccc}
q & b & r \\
\kappa & \beta & -\rho
\end{array} \right)
\left( \begin{array}{ccc}
r & c & p \\
\rho & \gamma & -\psi
\end{array} \right).
\label{clrotre}
\end{eqnarray}
\noindent Finally, the shelling $[3 \rightarrow 1]^{{\bf 3}}_{sh}$ is depicted 
on the top of FIG.3 
\begin{figure}[htb]
\leavevmode
\hspace{4.5cm} \epsfbox{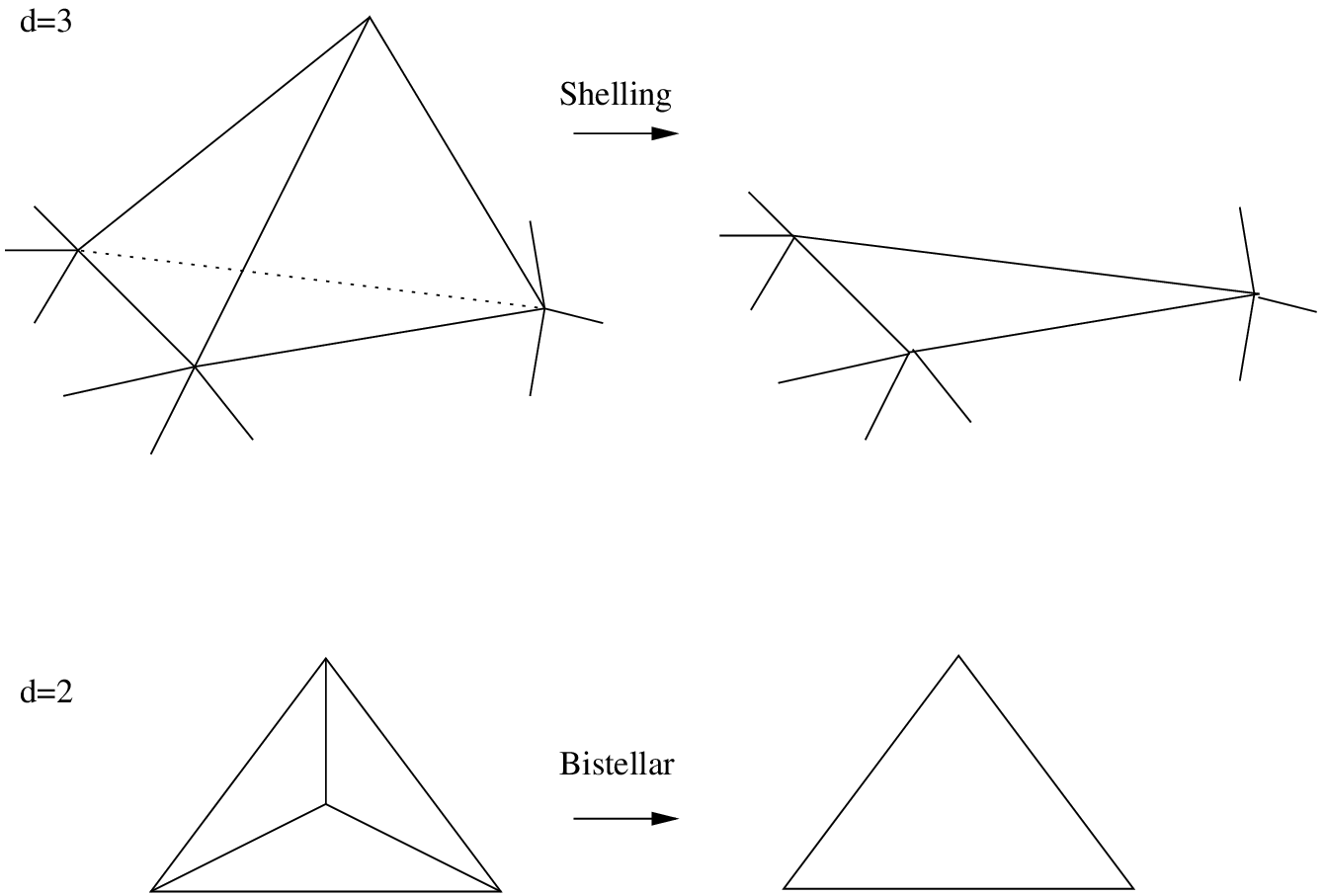}
\caption{ The shelling $[3 \rightarrow 1]^{{\bf 3}}_{sh}$ and the corresponding 
bistellar $[3 \rightarrow 1]^{{\bf 2}}_{bst}$.}
\label{shell2}
\end{figure}
and the associated identity reads\par
\begin{eqnarray}
\lefteqn{\Lambda(L)^{-1}\sum_{q\kappa ' ,p\psi ',r\rho '} (-1)^{-\psi ' -\kappa 
'-\rho '}
(-1)^{2(p+q+r)} (2p+1)(2r+1)(2q+1)\cdot}\nonumber\\
& & \cdot\left( \begin{array}{ccc}
a & p & q \\
\alpha & -\psi ' & \kappa '
\end{array} \right)
\left( \begin{array}{ccc}
b & q & r \\
\beta & -\kappa ' & \rho '
\end{array} \right) 
\left( \begin{array}{ccc}
c & r & p \\
\gamma & -\rho ' & \psi '
\end{array} \right)
 (-1)^{\Phi}\,
\left\{ \begin{array}{ccc}
a & b & c \\
r & p & q
\end{array}\right\}=\nonumber\\
& & =\,\left( \begin{array}{ccc}
b & a & c \\
\beta & \alpha & \gamma
\end{array} \right),
\label{clrouno}
\end{eqnarray}
\noindent where $\Lambda(L)$ is defined as in (\ref{clsum2}).\par
\noindent Notice that in each of the above identities we can read also the 
corresponding 
inverse shelling, namely the operation consisting in the attachment of a 
$3$-simplex 
to the suitable component(s) in $\partial T^3$, simply by exchanging the role of
internal and external labellings.\par
\noindent Comparing the above identities representing 
the elementary shellings and their inverse moves
with the expression given in (\ref{clsum2}), we see that 
the state sum $Z_{PR}$ $[(M^3, \partial M^3)]$ in (\ref{clsum1})  is formally
invariant both under (a finite number of) bistellar moves in the interior of 
$M^3$,
and under (a finite number of) elementary boundary operations. Following now 
\cite{pachner90} we are able to conclude that (\ref{clsum1}) is indeed an 
invariant of the
$PL$-structure (as well as a topological invariant, since we are dealing with 
$3$-dimensional $PL$-manifolds). \par

Since the structure of a local arrangment of $2$-simplices 
in the state sum (\ref{clsum1}) is naturally encoded in (\ref{clrodue}), 
(\ref{clrotre})
and (\ref{clrouno}), it turns out that a
state sum for a $2$-dimensional triangulation of a closed $PL$-manifold $M^2$ 
\begin{equation}
T^2(j;m,m') \longrightarrow M^2
\label{duetri}
\end{equation}
\noindent can be consistently defined if we require that\par
\begin{itemize}
\item each $2$-simplex $\sigma^2 \in T^2$ is associated with the following 
product of two Wigner symbols (a {\em double} $3jm$ symbol for short)\par
\end{itemize}
\begin{equation}
\sigma^2 \:\longleftrightarrow\: (-1)^{\sum_{s=1}^3 (m_s+m'_s)/2}
\left( \begin{array}{ccc}
j_1 & j_2 & j_3 \\
m_1 & m_2 & -m_3
\end{array}\right)
\left( \begin{array}{ccc}
j_1 & j_2 & j_3 \\
m'_1 & m'_2 & -m'_3
\end{array}\right),
\label{doublesim}
\end{equation}
\noindent where $\{m_s\}$ and $\{m'_s\}$ are two different sets of momentum 
projections associated with the
same angular momentum variables $\{j_s\}$, $-j \leq m_s, m'_s \leq j$ $\forall 
s=1,2,3$. 
The expression of the
state sum proposed in \cite{convegno} reads\par
\begin{eqnarray}
\lefteqn{Z[T^2(j;m,m')\rightarrow M^2; L] =}\hspace{.5in} \nonumber\\
& & =\Lambda(L)^{-N_0}\,\prod_{A=1}^{N_1} (2j_A+1) 
(-1)^{2j_A}(-1)^{-m_A-m'_A}\nonumber\\
& & \prod_{B=1}^{N_2} 
\left( \begin{array}{ccc}
j_1 & j_2 & j_3 \\
m_1 & m_2 & -m_3
\end{array}\right)_B
\left( \begin{array}{ccc}
j_1 & j_2 & j_3 \\
m'_1 & m'_2 & -m'_3
\end{array}\right)_B,
\label{bisum1}
\end{eqnarray}
\noindent where $N_0, N_1, N_2$ are the numbers of vertices, edges and triangles 
in $T^2$, respectively.
Summing over all of the admissible assignments of $\{j;m,m'\}$ we get\par
\begin{equation}
Z[M^2]\,=\,\lim_{L\rightarrow \infty}\:\sum_{\{T^2(j;m,m'), j \leq 
L\}}\:Z[T^2(j;m.m')\rightarrow M^2; L],
\label{bisum2}
\end{equation}
\noindent where the regularization is carried out according to the usual 
prescription.\par
\noindent The invariance of (\ref{bisum2}) 
is ensured as fas as the bistellar moves in $d=2$ can be implemented.
One of these move is expressed according to
\begin{eqnarray}
\lefteqn{\sum_q \sum_{\kappa,\kappa'}
(2q+1)(-1)^{2q}\,(-1)^{-\kappa -\kappa'}
\left( \begin{array}{ccc}
p & a & q \\
\psi & \alpha & -\kappa
\end{array}\right)
\left( \begin{array}{ccc}
q & b & r \\
\kappa & \beta & \rho
\end{array}\right) \cdot}\nonumber\\
& & \cdot \left( \begin{array}{ccc}
p & a & q \\
\psi' & \alpha' & -\kappa'
\end{array}\right)
\left( \begin{array}{ccc}
q & b & r \\
\kappa' & \beta' & \rho'
\end{array}\right)
\,=\,\sum_c 
\sum_{\gamma, \gamma'} (2c+1)\,(-1)^{2c}\,(-1)^{-\gamma -\gamma'}\cdot\nonumber\\
& & \cdot \left( \begin{array}{ccc}
a & b & c \\
\alpha & \beta & \gamma
\end{array}\right)
\left( \begin{array}{ccc}
r & p & c \\
\rho & \psi & -\gamma
\end{array}\right)
\left( \begin{array}{ccc}
a & b & c \\
\alpha' & \beta' & \gamma'
\end{array}\right)
\left( \begin{array}{ccc}
r & p & c \\
\rho' & \psi' & -\gamma'
\end{array}\right)
\label{dueflip}
\end{eqnarray}
\noindent and represents the so called {\em flip}, namely the bistellar move 
$[2 \rightarrow 2]^{{\bf 2}}_{bst}$, having taken into account the notation 
introduced in (\ref{dbst}) (refer to the bottom of FIG.1 for the corresponding 
picture).\par
\noindent The identity corresponding to the remaining moves, namely
$[1 \leftrightarrow 3]^{{\bf 2}}_{bst}$, reads\par
\begin{eqnarray}
\lefteqn{\sum_{q,r,p} 
(2q+1)(2r+1)(2p+1)\,(-1)^{2q+2r+2p}
\sum_{\kappa,\kappa'}\sum_{\rho,\rho'}\sum_{\psi,\psi'}
(-1)^{-\kappa -\kappa'}(-1)^{-\rho -\rho'} \cdot}\hspace{.5in}\nonumber\\
& & \cdot (-1)^{-\psi -\psi'} \left( \begin{array}{ccc}
p & a & q \\
\psi & \alpha & -\kappa
\end{array}\right)
\left( \begin{array}{ccc}
q & b & r \\
\kappa & \beta & -\rho
\end{array}\right)
\left( \begin{array}{ccc}
r & c & p \\
\rho & \gamma & -\psi
\end{array}\right)\cdot \nonumber\\
& & \cdot \left( \begin{array}{ccc}
p & a & q \\
\psi' & \alpha' & -\kappa'
\end{array}\right)
\left( \begin{array}{ccc}
q & b & r \\
\kappa' & \beta' & -\rho'
\end{array}\right)
\left( \begin{array}{ccc}
r & c & p \\
\rho' & \gamma' & -\psi'
\end{array}\right)\,=\nonumber\\
& & =\, \Lambda(L)^{-1}\,
\left( \begin{array}{ccc}
a & b & c \\
\alpha & \beta & \gamma
\end{array}\right)
\left( \begin{array}{ccc}
a & b & c \\
\alpha' & \beta' & \gamma'
\end{array}\right)
\label{duebary}
\end{eqnarray}
\noindent and these moves are depicted on the bottoms of FIG.2 and FIG.3.\par
\noindent As a matter of fact, the state sum given in (\ref{bisum1})
and (\ref{bisum2}) is formally invariant under (a finite number of) topological 
operations represented by (\ref{dueflip}) and
(\ref{duebary}). Thus, from Pachner's theorem proved in \cite{pachner87},
we conclude that it is a ($PL$) topological invariant. Its expression can be 
easily evaluated also in the $q$-case, providing us with a finite quantum 
invariant given by\par
\begin{equation}
Z^2_{\chi}(q) \equiv Z[M^2](q)\,=\,w^2_q\;w_q^{-2{\bf \chi}(M^2)},
\label{eulero}
\end{equation}
\noindent where ${\bf \chi}(M^2)$ is the Euler characteristic of the manifold 
$M^2$
and $w^2_q=-2k/(q-q^{-1})^2$ (see {\em Appendix B}).\par
\noindent In conclusion, the $2$-dimensional closed model generated by the 
$3$-dimensional model
with a non empty boundary is not trivial, being the only topological invariant 
which is significant for a closed surface in the present context.\par


\section{Extension of the Crane--Yetter model to\\
$(M^4, \partial M^4)$ 
and induced $3$-dimensional invariant}
In this section we revise first the results found in \cite{crane} 
(see also \cite{ooguri4d}
and \cite{ruth}) concerning the $q$-invariant $Z_{CY}[M^4](q)$ for a closed 
$PL$-manifold $M^4$. However, for the sake of simplicity, we limit ourselves 
to a detailed analysis of the $(q=1)$ case, and moreover we write down the 
expression
of the resulting $Z_{CY}[M^4]\doteq Z_{CY}[M^4](q)|_{q=1}$ in terms of the 
$3jm$ symbols appearing
in the definition of the $SU(2)$ $15j$ symbol of the second type (rather than 
using its
expression in terms of $6j$ coefficients). This last step turns out to be crucial
in order to define the new invariant $Z_{CY}[(M^4, \partial M^4)]$
for a $PL$-pair $(M^4, \partial M^4)$ and also in order to show that the state sum 
induced on $\partial M^4 \equiv M^3$ is indeed the topological invariant
$Z_{\chi}[M^3]$ (which is trivial in the present case since $M^3$ is a closed 
manifold).\par

Thus, consider a {\em multi--colored} triangulation of a given closed 
$PL$-manifold
$M^4$ denoting it by the map\par
\begin{equation}
T^4(j_{\sigma^2},\,J_{\sigma^3})\:\longrightarrow\:M^4,
\label{colquadri}
\end{equation}
\noindent where $j_{\sigma^2}$ is an $SU(2)$-coloring on the $2$-dimensional 
simplices
${\sigma^2}$ in $T^4$ and $J_{\sigma^3}$ is an $SU(2)$-coloring on tetrahedra 
${\sigma^3} \subset {\sigma^4}$ $\in T^4$ (recall that an ordering on the vertices
of each $4$-simplex ${\sigma^4}$ has to be chosen; however, the final 
expression of the state sum turns out to be independent of this choice). The 
consistency
in the assignment of the $\{j,J\}$ spin variables is ensured for a fixed ordering
if we require that\par
\begin{itemize}
\item each $3$-simplex ${\sigma^3}_a \subset {\sigma^4}$ ($a=1,2,\ldots,N_3$, 
$N_3$
being the number of $3$-simplices in $T^4$) must be associated, apart from a phase 
factor,
with a product of two $3jm$ symbols, namely\par
\begin{equation}
{\sigma^3}_a \:\longleftrightarrow \:
\left( \begin{array}{ccc}
j_1 & j_2 & J_a \\
m_1 & m_2 & m_a
\end{array}\right)
\left( \begin{array}{ccc}
j_3 & j_4 & J_a \\
m_3 & m_4 & -m_a
\end{array}\right);
\label{jJtresim}
\end{equation}
\item each $4$-simplex ${\sigma^4}\in T^4$ must be associated, apart from a phase 
factor,
with a summation of the product of ten suitable $3jm$ symbols (see below for its 
explicit expression),
giving rise to a $15j$ symbol of the second type which we represent for short 
as\par
\begin{equation}
\sigma^4 \:\longleftrightarrow \:
[J_a, J_b, J_c, J_d, J_e]_{\sigma^4},
\label{Jquasim}
\end{equation}
\noindent where $J_a,\ldots,J_e$ are labellings assigned to the five tetrahedra
${\sigma}_a,\ldots,{\sigma}_e \subset {\sigma}^4$.\par
\end{itemize}

\noindent Then we can define the following state sum\par
\begin{eqnarray}
\lefteqn{Z[T^4(j_{\sigma^2},J_{\sigma^3}) \rightarrow M^4 ; L] 
=}\hspace{.1cm}\nonumber \\
& & =\Lambda(L)^{N_0-N_1}\,\prod_{\sigma^2 \in T^4} 
(-1)^{2j_{\sigma^2}}(2j_{\sigma^2}+1)
\prod_{\sigma^3 \in T^4} (-1)^{2J_{\sigma^3}}(2J_{\sigma^3}+1) \cdot\nonumber\\
& & \cdot \prod_{\sigma^4 \in T^4}\; 
[J_a,J_b,J_c,J_d,J_e]_{\sigma^4},
\label{cysum}
\end{eqnarray}
\noindent where $N_0,N_1$ are the number of vertices and edges in $T^4$, 
respectively.
The $15j$ symbol associated with each $4$-simplex is given explicitly by\par
\begin{eqnarray}
\lefteqn{ [J_a, J_b, J_c, J_d, J_e]_{\sigma^4} \doteq \{15j\}_{\sigma^4}(J)
=}\hspace{.5cm}\nonumber \\
& & =\sum_{m}(-1)^{\sum m}\,
\left(\begin{array}{ccc}
j_1 & j_2 & J_a \\
m_1 & m_2 & m_a
\end{array}\right)
\left(\begin{array}{ccc}
j_3 & j_4 & J_a \\
m_3 & m_4 & -m_a
\end{array}\right)
\left(\begin{array}{ccc}
j_5 & j_6 & J_b \\
m_5 & m_6 & m_b
\end{array}\right)\cdot\nonumber\\
& & \cdot \left(\begin{array}{ccc}
j_3 & j_7 & J_b \\
-m_3 & m_7 & -m_b
\end{array}\right)
\left(\begin{array}{ccc}
j_5 & j_8 & J_c \\
-m_5 & m_8 & m_c
\end{array}\right)
\left(\begin{array}{ccc}
j_1 & j_9 & J_c \\
-m_1 & m_9 & -m_c
\end{array}\right)\cdot\nonumber\\
& & \cdot\left(\begin{array}{ccc}
j_6 & j_{10} & J_d \\
-m_6 & m_{10} & -m_d
\end{array}\right)
\left(\begin{array}{ccc}
j_2 & j_8 & J_d \\
-m_2 & -m_8 & -m_d
\end{array}\right)
\left(\begin{array}{ccc}
j_7 & j_{10} & J_e \\
-m_7 & -m_{10} & -m_e
\end{array}\right)\cdot\nonumber\\
& & \cdot \left(\begin{array}{ccc}
j_4 & j_9 & J_e \\
-m_4 & -m_9 & -m_e
\end{array}\right),
\label{15j}
\end{eqnarray}
\noindent where the summation is extendend to all admissible values of the $m$ 
variables,
and the planar diagram corresponding to the symbol is sketched in FIG.4.\par
\begin{figure}[htb]
\leavevmode
\hspace{4cm} \epsfbox{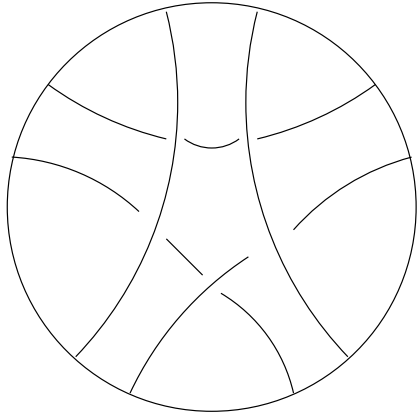}
\caption{ Diagrammatic representation of the 15j symbol of the second type.}
\label{15ja}
\end{figure}
\noindent It can be shown (see \cite{crane}, \cite{roberts}) that the
expression\par
\begin{equation}
Z_{CY}[M^4]\,=\,
\lim_{L\rightarrow \infty}\:\sum_{
\left\{\begin{array}{c}
T^4(j,J)\\
j,J \leq L
\end{array}\right\}}
Z[T^4(j_{\sigma^2},J_{\sigma^3})\rightarrow M^4; L]
\label{cyinv}
\end{equation}
\noindent is formally a $PL$-invariant, and that its value is given
by $\Lambda(L)^{\chi (M^4)/2}$ $K^{\sigma(M^4)}$, where $\chi (M^4)$, 
$\sigma(M^4)$ are
the Euler characteristic and the signature of $M^4$ respectively, and $K$ is a 
constant.\par
The former state sum can be generalized to the case of a compact $4$-dimensional
$PL$-pair by considering the map\par
\begin{equation}
(T^4(j_{\sigma^2},J_{\sigma^3}), \partial 
T^4(j'_{\sigma^2},J'_{\sigma^3};m_{\sigma^2},
m_{\sigma^3}))
\,\longrightarrow\,(M^4,\partial M^4),
\label{bouquadri}
\end{equation}
\noindent where $\{j_{\sigma^2},J_{\sigma^3}\}$ denotes the entire set of spin 
variables,
ranging from $1$ to $N_2$ and from $1$ to $N_3$, respectively. The subset 
$\{j'_{\sigma^2},J'_{\sigma^3}\}$ $\subset$ $\{j_{\sigma^2},J_{\sigma^3}\}$ 
contains the
colorings of the subsimplices in $\partial T^4$, the corresponding magnetic 
numbers of
which can be collectively denoted by $m \equiv$ $\{m_{\sigma^2},m_{\sigma^3}\}$ 
as far as no confusion arises. The assignment of
the above variables turns out to be consistent if we agree with the statements in
(\ref{Jquasim})(for all $\sigma^4$ $\in (T^4,\partial T^4)$, taking into 
account the fact that some of the $J$ labels may become $J'$ for those 
$4$-simplices 
which have component(s) in $\partial T^4$)and with
(\ref{jJtresim})(for $3$-simplices in the interior of the triangulation).
Moreover, we require that\par
\begin{itemize}
\item each $3$-simplex ${\sigma^3}_a \subset {\sigma^4}$ lying in the boundary
$\partial T^4$ must be associated, apart from a phase factor,
with the following product of $3jm$ symbols\par
\begin{equation}
{\sigma^3}_a \:\longleftrightarrow \:
\left( \begin{array}{ccc}
j'_1 & j'_2 & J'_a \\
m_1 & m_2 & m_a
\end{array}\right)
\left( \begin{array}{ccc}
j'_3 & j'_4 & J'_a \\
m_3 & m_4 & -m_a
\end{array}\right).
\label{boutresim}
\end{equation}
\end{itemize}

\noindent With these premises, we consider now the following expression\par 
\begin{eqnarray}
\lefteqn{Z[(M^4,\partial M^4)]=}\hspace{-1cm}\nonumber\\
& & =\,\lim_{L\rightarrow \infty}\,\sum_{
\left\{\begin{array}{c}
(T^4,\partial T^4)\\
j,J,j',J';\\
m \leq L
\end{array}\right\}}Z[
(T^4(j,J), \partial T^4(j',J';m)) \rightarrow
(M^4,\partial M^4);L],
\label{quadrinv}
\end{eqnarray}
\noindent where we have used a shorthand notation instead of (\ref{bouquadri}),
and where\par
\begin{eqnarray}
\lefteqn{Z[(T^4(j,J), \partial T^4(j',J',;m)] \rightarrow (M^4,\partial M^4); L] 
=}\hspace{.2cm}\nonumber \\
& & =\Lambda(L)^{N_0-N_1}\,\prod_{all\, \sigma^2} 
(-1)^{2j_{\sigma^2}}(2j_{\sigma^2}+1)
\prod_{all\, \sigma^3} (-1)^{2J_{\sigma^3}}(2J_{\sigma^3}+1) \cdot\nonumber\\
& & \cdot \prod_{all\, \sigma^4} \{15j\}_{\sigma^4}(J,J')
\prod_{\sigma^3 \in \partial T^4}(-1)^{\sum m_{j'}/2+ \sum m_{J'}}
\left( \begin{array}{ccc}
j'_1 & j'_2 & J'_a \\
m_1 & m_2 & m_a
\end{array}\right)\cdot \nonumber\\
& & \cdot \left( \begin{array}{ccc}
j'_3 & j'_4 & J'_a \\
m_3 & m_4 & -m_a
\end{array}\right).
\label{quadrisum}
\end{eqnarray}
\noindent Here we introduced explicitly in the phase factors $m_{j'}$ and $m_{J'}$ 
to denote magnetic numbers 
corresponding to different kinds of spin variables on the boundary.\par

As discussed in {\em Section 2}, for a $PL$-pair in dimension $d$ there exist $d$
different types of elementary shellings (and $d$ inverse shellings), parametrized
by the number $n_{d-1}$ of faces in a boundary $d$-simplex according to 
(\ref{dshel}). 
Thus in the present case we are dealing with four different shellings, 
$n_3=1,2,3,4$ 
being the number of tetrahedra in $\partial T^4$ which are going to disappear
(together with the underlying $4$-simplex), respectively.
The diagrammatic representations of the moves $[1\rightarrow 4]_{sh}^{{\bf 4}}$,
$[2\rightarrow 3]_{sh}^{{\bf 4}}$, $[3\rightarrow 2]_{sh}^{{\bf 4}}$,
$[4\rightarrow 1]_{sh}^{{\bf 4}}$ are displayed in FIG.5, 
\begin{figure}[htb]
\leavevmode
\hspace{4.5cm} \epsfbox{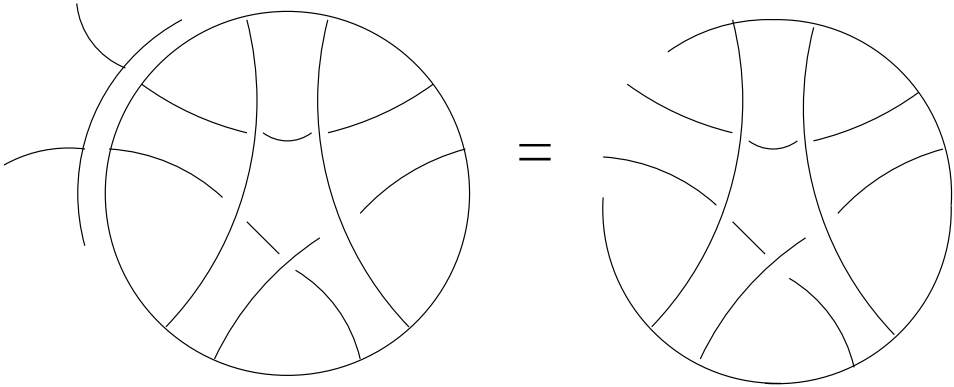}
\caption{ Diagrammatic representation of the move 
$[1\rightarrow 4]_{sh}^{{\bf 4}}$.}
\label{14shel}
\end{figure}
FIG.6,
\begin{figure}[htb]
\leavevmode
\hspace{4.5cm} \epsfbox{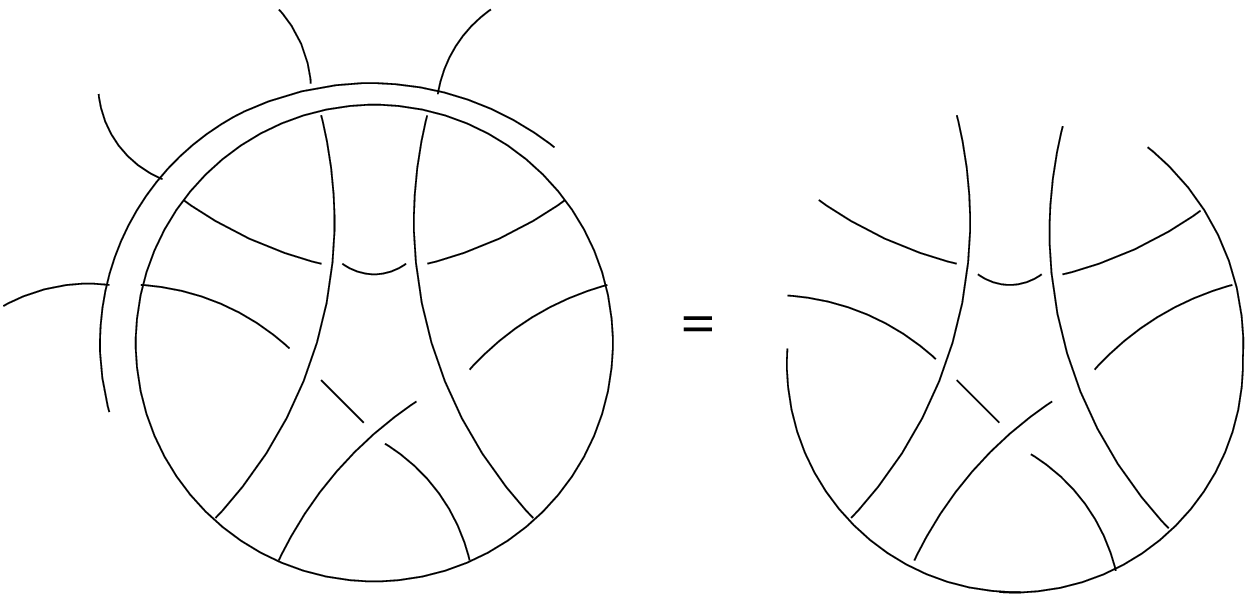}
\caption{ Diagrammatic representation of the move 
$[2\rightarrow 3]_{sh}^{{\bf 4}}$.}
\label{23shel}
\end{figure}
FIG.7, 
\begin{figure}[htb]
\leavevmode
\hspace{2.5cm} \epsfbox{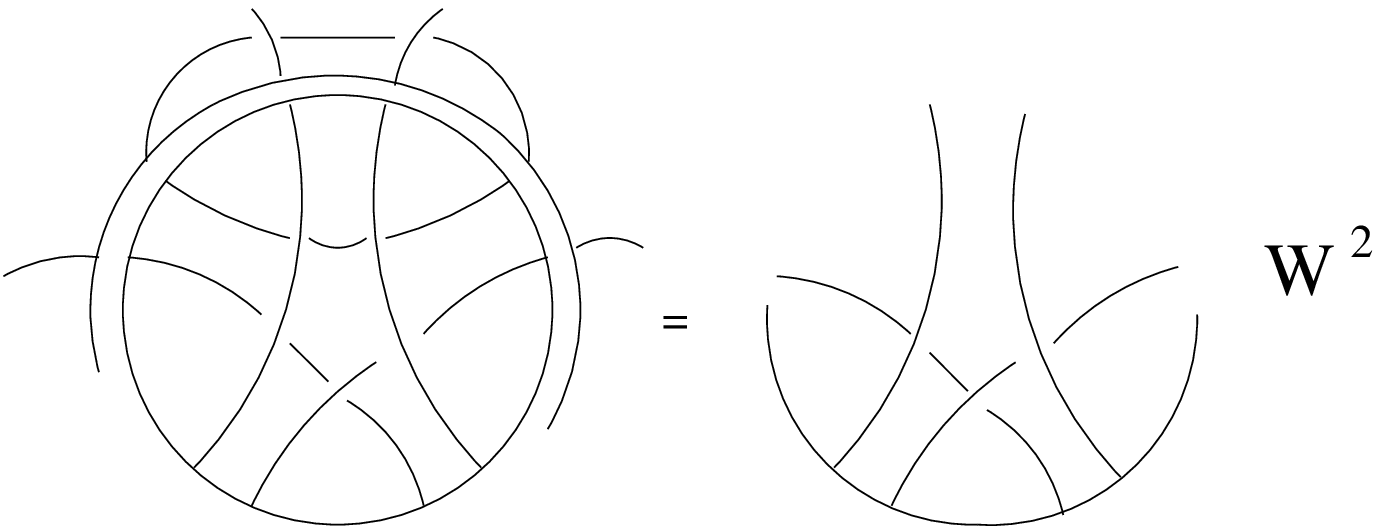}
\caption{ Diagrammatic representation of the move 
$[3\rightarrow 2]_{sh}^{{\bf 4}}$.}
\label{32shel}
\end{figure}
FIG.8, 
\begin{figure}[htb]
\leavevmode
\hspace{4.5cm} \epsfbox{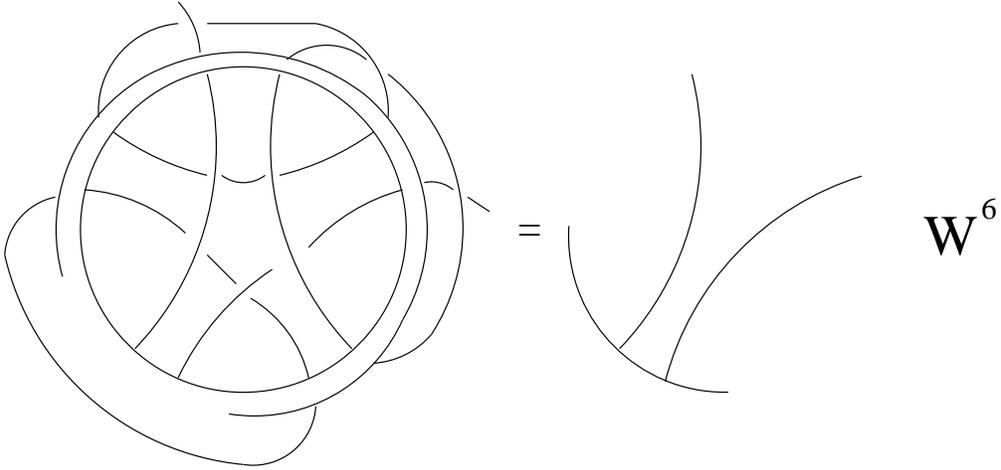}
\caption{ Diagrammatic representation of the move 
$[4\rightarrow 1]_{sh}^{{\bf 4}}$.}
\label{41shel}
\end{figure}
respectively, where we have made use of diagrammatical
relations to handle products of $3jm$ symbols (see FIG.9 
\begin{figure}[htb]
\leavevmode
\hspace{2.5cm} \epsfbox{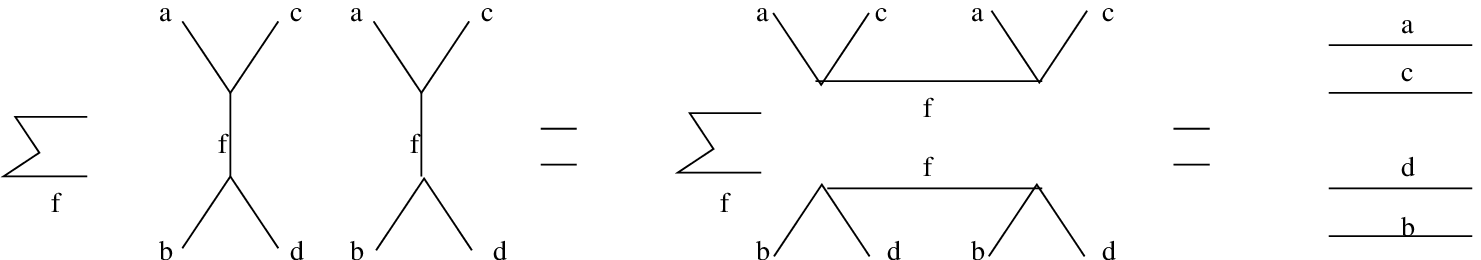}
\caption{ Relation involving the product of two couple of 3jm symbols.}
\label{rel1}
\end{figure}
and FIG.10).
\begin{figure}[htb]
\leavevmode
\hspace{5cm} \epsfbox{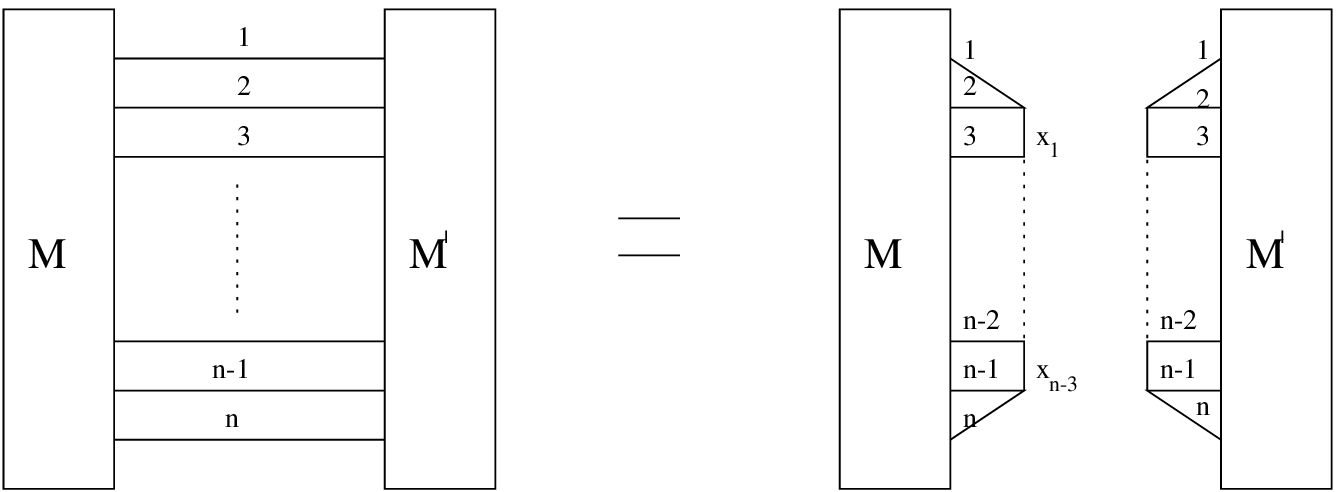}
\caption{ Diagrammatical representation of the general relation 
involving the product of two
quantities containing many 3jm symbols; 
$M$ and $M'$ represent the portions of the diagram leaved unchanged by the
relation.}
\label{relg}
\end{figure} 
The explicit expressions of the algebraic identities associated
with elementary shellings
are collected in {\em Appendix A}.\par 
\noindent According to these remarks, $Z[(M^4, \partial M^4)]$ 
in (\ref{quadrinv}) turns out to be
formally equivalent under the action  of a finite set of the above operations 
and their inverse moves and thus, owing to the theorem proved in \cite{pachner90}, 
it defines an invariant of the $PL$-structure. Its $q$-deformed counterpart,
$Z[(M^4,\partial M^4)](q)$, can be worked out according to the prescriprion
given in {\em Appendix B} and represents a well--defined quantum invariant.\par
\noindent We may notice also that, since (\ref{quadrinv}) reduces to (\ref{cyinv})
when $\partial M^4=\emptyset$, $Z[(M^4, \partial M^4)]$ is invariant under 
bistellar moves performed in $int(T^4)$ too. However, we will show in 
{\em Section 6} that the equivalence of our $d$-dimensional $Z[(M^d, \partial 
M^d)]$ 
under elementary shellings implies (in a non trivial way) the invariance under 
bistellar moves of the state sum induced by setting $\partial M^d =\emptyset$.\par
Looking now at the local arrangement of $3$-simplices in the state sum 
(\ref{quadrisum}),
it turns out that an {\em induced} $3$-dimensional state sum for a colored closed 
triangulation\par
\begin{equation}
T^3(j_{\sigma^2},J_{\sigma^3}; m_{\sigma^2},
\mu_{\sigma^2}; m_{\sigma^3},\mu_{\sigma^3})
\,\longrightarrow\,M^3
\label{tridtri}
\end{equation}
\noindent can be defined by associating with each $3$-simplex the following sum
of products of Wigner symbols\par
\begin{eqnarray}
\lefteqn{
\sigma^3_a \longleftrightarrow
\sum_{m_a,\mu_a}(-1)^{m_a+\mu_a}
\left( \begin{array}{ccc}
j_1 & j_2 & J_a \\
m_1 & m_2 & m_a
\end{array}\right)
\left( \begin{array}{ccc}
j_3 & j_4 & J_a \\
m_3 & m_4 & -m_a
\end{array}\right)\cdot}\hspace{1cm}\nonumber\\
& & \cdot \left( \begin{array}{ccc}
j_1 & j_2 & J_a \\
\mu_1 & \mu_2 & \mu_a
\end{array}\right)
\left( \begin{array}{ccc}
j_3 & j_4 & J_a \\
\mu_3 & \mu_4 & -\mu_a
\end{array}\right)\doteq\nonumber\\
& & \doteq
 \left( \begin{array}{ccccc}
j_1 & j_2 & J_a & j_3 & j_4\\
m_1 & m_2 & & m_3 & m_4\\
\mu_1 & \mu_2 & & \mu_3 & \mu_4
\end{array}\right)_{\sigma^3_a}.
\label{doubtresim}
\end{eqnarray}
\noindent Here all magnetic numbers $\{m,\mu\}$ have their natural ranges of 
variation
with respect to the corresponding $\{j,J\}$ and a shorthand notation for
the product of symbols has been introduced. According to (\ref{doubtresim}), we 
can
define the following state sum\par
\begin{eqnarray}
\lefteqn{Z[T^3(j,J;m,\mu) \rightarrow M^3) ; L] =}\hspace{.5cm}\nonumber \\
& & =\Lambda(L)^{N_0-N_1}\,\prod_{all\,\sigma^2 } (2j_{\sigma^2}+1)
\prod_{all\,\sigma^3_a}(2J_{\sigma^3_a}+1)(-1)^{\sum (J+j)} \cdot\nonumber\\
& & \cdot \left( \begin{array}{ccccc}
j_1 & j_2 & J_a & j_3 & j_4\\
m_1 & m_2 & & m_3 & m_4\\
\mu_1 & \mu_2 & & \mu_3 & \mu_4
\end{array}\right)_{\sigma^3_a},
\label{tridsum}
\end{eqnarray}
\noindent which gives rise to the expression
\begin{equation}
Z[M^3]=
\lim_{L\rightarrow \infty}\,\sum_{
\left\{\begin{array}{c}
T^3(j,J,m,\mu)\\
j,J,m,\mu \leq L
\end{array}\right\}}
Z[T^3(j,J;m,\mu)
\rightarrow M^3;L].
\label{tridinv}
\end{equation}
\noindent At this point we should implement those 
topological operations which are suitable in the present context, 
namely the bistellar moves 
$[n_3 \rightarrow 5-n_3]^{{\bf 3}}_{bst}$, $n_3=1,2,3,4$. However,
we may provide a straightforward proof that $Z[M^3]$ in (\ref{tridinv}) is
related to the Euler characteristic of $M^3$. The combination of symbols in 
each $\sigma^3_a$ ({\em cfr.} (\ref{doubtresim}))
can be interpreted (putting the spin variables on the same foot) as a contribution
of two triangles joined along $J_a$ in a triangulation $S^2$ of a $2$-dimensional
surface uniquely associated with the given $T^3$ in (\ref{tridsum}). Then, by 
comparing
such a structure with (\ref{doublesim}) and with $Z^2_{\chi}$ in (\ref{bisum2}), 
we see that the contribution
for a finite $L$ of the sums over all $1$-simplices (from $1$ to $N_1$) and over
all $2$-simplices (from $1$ to $N_2$) in $S^2$ amounts to $w_L^2 \cdot 
w_L^{2(N_1-N_2)}$,
where we set $\Lambda(L)^{-1}\equiv w_L^{-2}$. By considering again 
(\ref{doubtresim})
and the fact that we are dealing with manifolds, it turns out that
the number of $1$-simplices in $S^2$ is related to $N_2$ and $N_3$ in $T^3$ 
by $N_1(S^2)=(N_2+N_3)(T^3)$, and also that $N_2(S^2)=2N_3(T^3)$. Thus, for each 
finite
value of $L$, we would obtain $Z[M^3;L]$= $w_L^{2+2\chi(M^3)}$.
The regularized version of this results reads\par
\begin{equation}
Z_\chi[M^3](q)\,=\,w_q^{2[1+\chi(M^3)]}\equiv w_q^2,
\label{chitre}
\end{equation}
\noindent where we have taken into account the fact that the Euler characteristic
vanishes for any $3$-dimensional closed manifold.\par

\section{Invariants of closed $M^d$ from colorings of \\
$(d-1)$-simplices}

An alternative way of defining the state sums which give rise to $Z^2_{\chi}$
and $Z^3_{\chi}$ comes out when we consider the relationships 
between integrals of products of Wigner $D$-functions and suitable products of 
$3jm$ symbols (see {\em e.g.} \cite{russi}). Picking up the $1$-skeleton of the 
dual 
complex of a given triangulation (either $T^2$ or $T^3$) we can assign in a
consistent way a $D^j_{m\mu}(R)$, $R\in SU(2)$, to each edge incident on
the vertices of such graphs (obviously the vertices are $3$-valent 
in $d=2$ and $4$-valent
in $d=3$). In this framework the role of the magnetic quantum numbers $m,\mu$ 
is made manifest by introducing the fat graph associated with each one of the 
former 
graphs: thus any edge acquires two sets of $SU(2)$-colorings, namely $\{j,m\}$ and
$\{j,\mu\}$. The next step consists in performing an integral over 
the $R$-variables of the product of the $D$-functions associated with the
legs of the graph incident on each vertex. By collecting the
terms generated by all vertices, we would get exactly 
the products of double $3jm$ symbols (with the correct phase factors) which
appear in the expressions of both $Z^2_{\chi}$ and $Z^3_{\chi}$. 
The formal calculation can be easily translated in the diagrammatic language as
shown in FIG.11.\par
\begin{figure}[htb]
\leavevmode
\hspace{3.5cm} \epsfbox{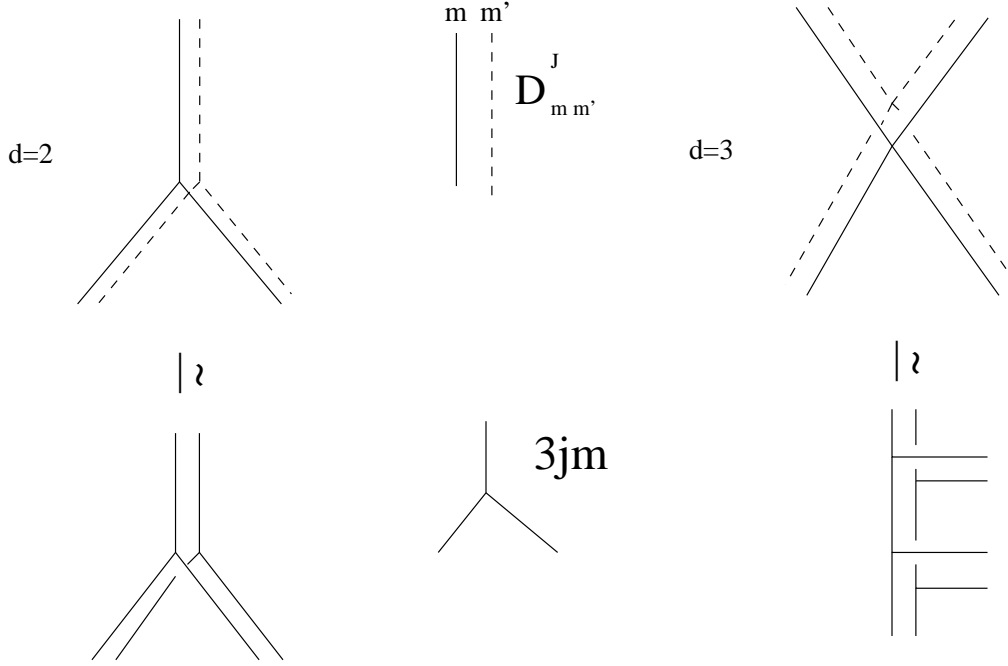}
\caption{ The graph representing the 1-skeleton of the dual lattice of each 2
and 3-dimensional simplex, with a $D$ function associated with each edge, and the
derived  elementary diagrams occurring in $Z^2_{\chi}$ and $Z^3_{\chi}$ in
terms of 3jm symbols.}
\label{zx23}
\end{figure}
The above procedure can be generalized to any dimension $d$. In particular, given
any triangulation $T^d$ of a closed $PL$-manifold $M^d$, we focus our 
attention on its dual $1$-skeleton, which is a $(d+1)$-valent graph $\Gamma$. 
As
before, we are going to associate a $D$-function with each {\em fat} edge
incident on each vertex. This amounts to require that consistent 
$SU(2)$-colorings on $\Gamma$ generated by the fat graph are achieved if 
we consider the assignment\par
\begin{equation}
({\{j\}};m,\mu)\;\longrightarrow\;\Gamma \,\subset {\tilde T^d},
\label{dichitri}
\end{equation}
\noindent where ${\tilde T^d}$ is the dual complex associated with $T^d$.
The pair of magnetic numbers for each ${j}$ variable
refers to what we said before, namely we will have a double symbol, each 
component of which displays the same ${\{j\}}$ but different $m$, $\mu$, 
for each elementary configuration
of the dual graph.\par
\noindent Moreover, it should be clear that the assignments in (\ref{dichitri})
can be thought of as a coloring on the $(d-1)$-skeleton of the original 
triangulation. 
According to this prescription, we can now define the following
limit of admissible sums of configurations written in terms of ${\{j\}}$
variables\par
\begin{eqnarray}
\lefteqn{Z_{\chi}[M^d]\:=\lim_{L\rightarrow \infty}\:\sum_{\left\{\begin{array}{c}
T^d({\{j\}};m,\mu)\\
{\{j\}},m,\mu\leq L
\end{array}\right\}}
w_L^{(-1)^{(d-1)}\Psi}\cdot}\hspace{.1cm}\nonumber\\
& & \cdot \prod_{all\,\sigma^{d-1}}\,(-1)^{2j_{\sigma^{d-1}}}
(2j_{\sigma^{d-1}}+1)\,
\left(\int \prod_{\sigma^{d-1}\subset
\sigma^{d}}
D^{{j_{\sigma^{d-1}}}}_{m\mu}(R)dR\right)_{\sigma^{d}},
\label{dichisum1}
\end{eqnarray}
\noindent where $w_L^2=\Lambda(L)$ and $\Psi=2(N_0-N_1+\ldots+(-1)^{d-2}N_{d-2})$, 
($N_0,N_1,N_2,\ldots$)
are the numbers of ($0,1,2,\ldots$)-simplices in $T^d$. The 
range of variation
of each $m,\mu$ with respect to the corresponding $j$ is the usual one, and  
the summations over the magnetic numbers act as
glueing operations among labelled $d$-dimplices.\par
\noindent Now we can exploit the relationships between integrals of products
of $D$-functions and suitable combinations of Wigner symbols: this amounts
to recognize that the integration in (\ref{dichisum1}) 
can be recasted as\par 
\begin{eqnarray}
\lefteqn{\sum_{\{J\}}Z[\sigma^d(\{j\},\{J\};m,\mu;{\cal M},\nu)\,\rightarrow M^d;L]\,
\equiv}\hspace{.2cm}\nonumber\\
& & \equiv \sum_{\{J\}}\prod_{k=1}^{d-3}(-1)^{2J_k}(2J_k+1)\sum_{{\cal M},\nu}
(-1)^{\sum{\cal M}+\sum \nu}
\left( \begin{array}{ccc}
j_1 & j_2 & J_1 \\
m_1 & m_2 & {\cal M}_1
\end{array}\right)\cdot\nonumber\\
& & \left( \begin{array}{ccc}
J_1 & j_3 & J_2 \\
-{\cal M}_1 & m_3 & {\cal M}_2
\end{array}\right)\cdot
\left(\begin{array}{ccc}
j_1 & j_2 & J_1 \\
\mu_1 & \mu_2 & \nu_1
\end{array}\right)
\left( \begin{array}{ccc}
J_1 & j_3 & J_2 \\
-\nu_1 & \mu_3 & \nu_2
\end{array}\right)\cdot \cdot \cdot \nonumber\\
& & \cdot \cdot \cdot \left( \begin{array}{ccc}
J_{{d}-4} & j_{{d}-2} & J_{{d}-3} \\
-{\cal M}_{{d}-4} & m_{{d}-2} & {\cal M}_{{d}-3}
\end{array}\right)
\left( \begin{array}{ccc}
J_{{d}-3} & j_{{d}-1} & j_{{d}} \\
-{\cal M}_{{d}-3} & m_{{d}-1} & m_{{d}}
\end{array}\right)\cdot\nonumber\\
& & \cdot \left( \begin{array}{ccc}
J_{{d}-4} & j_{{d}-2} & J_{{d}-3} \\
-\nu_{{d}-4} & \mu_{{d}-2} & \nu_{{d}-3}
\end{array}\right)
\left( \begin{array}{ccc}
J_{{d}-3} & j_{{d}-1} & j_{{d}} \\
-\nu_{{d}-3} & \mu_{{d}-1} & \mu_{{d}}
\end{array}\right).
\label{dichisum2}
\end{eqnarray}
\noindent 
Here we put in evidence, besides the ${\{j\}}$ colorings, also the set $\{J\}$, 
the role of which is similar to what
we found {\em e.g.} in (\ref{doubtresim}), namely $J$ variables come out to
be associated with an internal glueing between 
the two sub--symbols of a double symbol.
Moreover, we denoted by ${\cal M},\nu$ the two sets of magnetic numbers
associated with each $J$ and by $m,\mu$ those associated with each $j$, 
respectively. The diagrammatic counterpart of the procedure described above is
shown in FIG.12, 
\begin{figure}[htb]
\leavevmode
\hspace{1.5cm} \epsfbox{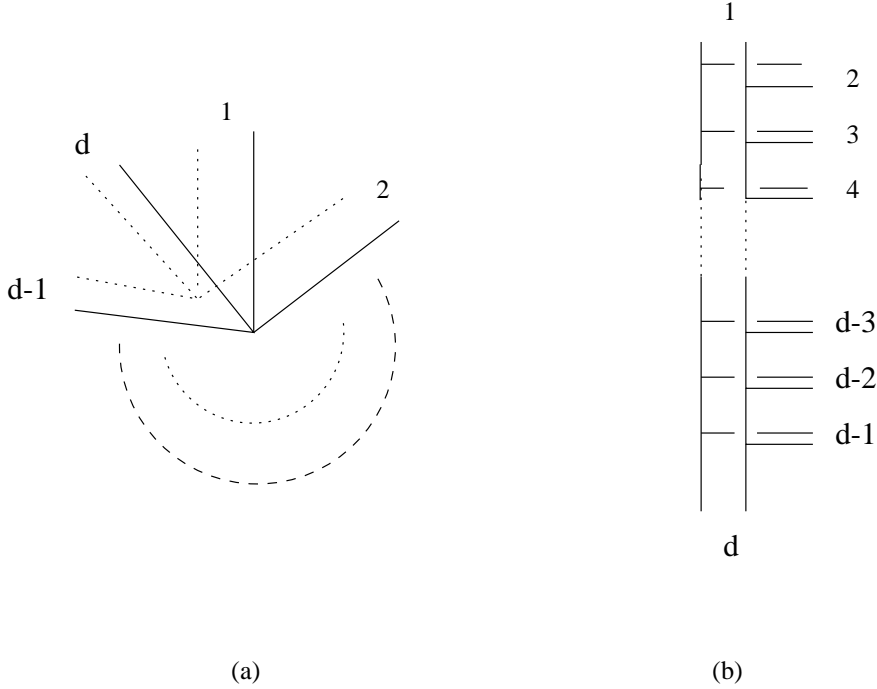}
\caption{  The graph representing the 1-skeleton of the dual lattice of each 
d-dimensional simplex, with a $D$ function associated with each edge, and the
derived  elementary diagrams occurring in $Z^d_{\chi}$ in
terms of 3jm symbols.}
\label{4nzx}
\end{figure}
where for simplicity just one of the possible coupling
schemes is considered (the other ones giving 
equivalent analytical expressions).\par 
\noindent Thus the limit in (\ref{dichisum1}) can be rewritten as
\begin{eqnarray}
\lefteqn{Z[M^d]=\lim_{L\rightarrow \infty}
\sum_{\left\{\begin{array}{c}
T^d(\{j\},\{ J\};m,\mu)\\
all \:j,m \leq L
\end{array}\right\}}w_L^{(-1)^{(d-1)}\Psi}\cdot}\hspace{.5cm}\nonumber\\
& & \prod_{all\,\sigma^{d-1}}\,(-1)^{2j_{\sigma^{d-1}}}
(2j_{\sigma^{d-1}}+1)\,
\prod_{k=1}^{d-3}(-1)^{2J_k}(2J_k+1)
\sum_{{\cal M},\nu}(-1)^{\sum{\cal M}+\sum\nu}\cdot\nonumber\\
& & \cdot Z[\sigma^d(\{j\},\{J\};m,\mu;{\cal M},\nu)\,\rightarrow M^d;L],
\label{dichisum3}
\end{eqnarray}
\noindent where we used the shorthand notation defined in (\ref{dichisum2}).\par
\noindent An explicit calculation which involves the auxiliary surface $S^2$ ({\em cfr.}
the discussion at the end of {\em Section 3}), where now we get $(d-2)$
triangles associated with each $\sigma^{d}$ in the original triangulation $T^d$, 
yields the regularized result\par
\begin{equation}
Z[M^d](q)\;=\;w_q^{2[1+(-1)^{(d-1)} \chi(M^d)]},
\label{deulero}
\end{equation}   
\noindent where $\chi(M^d)$ is the Euler characteristic of $M^d$.\par 
\noindent On passing, we may note that the counterpart of (\ref{deulero})
in the continuum approach would be the $d$-dimensional topological
field theory with a supersymmetric action given {\em e.g.} in \cite{eulero}.\par

\section{Invariants of $(M^d,\partial M^d)$ induced by $Z_{\chi}^{d-1}$}
According to the program outlined in {\em steps 1),2)} of the
introduction, we build up in the following a state sum
for a pair $(T^d,\partial T^d)$ induced by examinig the expression of 
$Z_{\chi}[\partial M^d \equiv
M^{d-1}]$, with $Z_{\chi}[M^{d-1}]$ given by (\ref{dichisum2}) of 
{\em Section 4}. The second part of the present section will be devoted to
the proof that such state sums are actually independent of the triangulation
chosen, and thus the invariant $Z[(M^d,\partial M^d)]$ is well defined in
any dimension $d$.\par
\noindent We learnt from the procedure followed both in {\em Section 2} and
in {\em Section 3} that once we give the {\em double} symbol associated with the
$(d-1)$-simplex in a given closed $T^{d-1}$, we can recover the contribution from 
a single 
$d$-simplex in $(T^d,\partial T^d\equiv T^{d-1})$ by taking first the 
{\em squared root} of the symbol itself. 
Then the recoupling symbol to be associated with the $d$-dimensional 
fundamental block is obtained by summing over the free $m$ entries the product
of $(d+1)$-contributions from its faces, with suitable $j$ labels.
Thus in $d=2$ we used the symbol given in (\ref{doublesim}) as a
fundamental block in the state sum $Z[T^2(j;m,m')\rightarrow M^2;L]$
(see (\ref{bisum1})), while in $d=3$ the symbol given in (\ref{doubtresim})
was associated with each $3$-simplex of $T^3(j,J;m,\mu)\rightarrow M^3$
(see (\ref{tridsum})): in both cases the double symbol looks like a square
of some sub--symbol. Taking the squared root means that we pick up just one
of the sub--symbols ({\em e.g.} either the Wigner symbol with $m$ entries in
(\ref{doublesim}) or the product of two Wigner symbols with $m$ entries
in (\ref{doubtresim})). By summing over all $m$ entries the product
of four sub--symbols with suitable $j$ labellings and phase factors in $d=3$ 
we get the expression of the $6j$ symbol which enters $Z_{PR}[M^3]$. In turn, a
summation over $m$ variables of the product of five sub--symbols 
with suitable $j$ labellings and phase factors in $d=4$ reproduces the 
$15j$ symbol which represents the fundamental block in
$Z_{CY}[M^4]$ (see (\ref{15j})). Notice also that the procedure works with
$PL$-pairs as well: we simply associate one of the former sub--symbol
(with an $m'$ coloring, say) with each $(d-1)$-simplex in $\partial T^d$.\par
The above remarks suggest how an algorithmic procedure for generating 
$Z[(M^d,$ $\partial M^d)]$ from $Z_{\chi}[M^{d-1}]$ could be actually established.
To this end, we consider first the structure of the double symbol associated 
with the fundamental block in (\ref{dichisum2}), written for a closed 
triangulation $T^{d-1}$. The corresponding planar graph is shown in FIG.13
\begin{figure}[htb]
\leavevmode
\hspace{4.5cm} \epsfbox{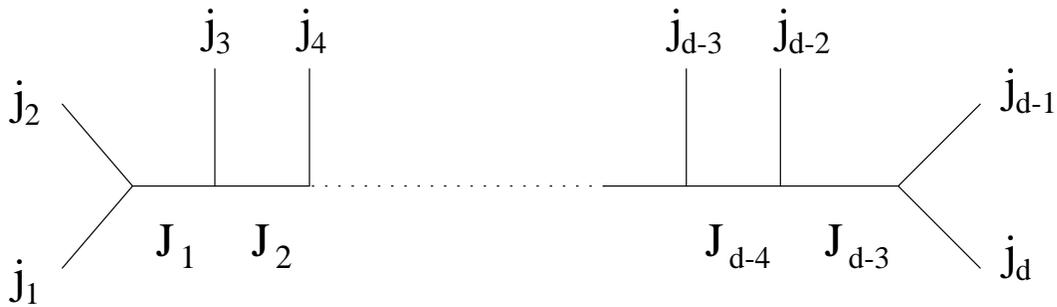}
\caption{ Diagram corresponding to the fundamental block 
occuring in the recoupling symbol 
associated with the 
d-dimensional simplex.}
\label{fonbloc}
\end{figure}
(compare also FIG.12): the diagram includes $d$ external legs, representing
the faces of the $(d-1)$-simplex, and $(d-3)$ internal edges, the colorings
of which are associated with the $(d-1)$-simplex itself, as explained below. 
The fundamental $d$-dimensional block which will enter the state sum is obtained 
by
assembling $(d+1)$ simplices of dimension $(d-1)$ along their
$(d-2)$-dimensional faces. Such a procedure can be described in some
detail as follows. We first assign an overall ordering to the set of 
$(d-1)$-simplices of $\sigma^{d}$, namely we introduce $a=1,2,\ldots,{d+1}$. 
Then we denote by $j_1^a,j_2^a,\ldots,j_d^a$ the $j$ labels of the external legs
of the graph associated with the $a$-th $(d-1)$-simplex $\sigma_a^{d-1}$ (the 
dimensionality of such colored faces is $(d-2)$). From a topological point of
view, we are going to join a suitable number of other colored $(d-1)$-simplices 
along the faces of the chosen $\sigma_a^{d-1}$ according to the rule:
$\sigma_{a+1}^{d-1}$ $\cap$ $\sigma_a^{d-1}=j_d^a$, 
$\sigma_{a-1}^{d-1}$ $\cap$ $\sigma_a^{d-1}=j_1^a$, $\ldots$,
$\sigma_{a-i}^{d-1}$ $\cap$ $\sigma_a^{d-1}=j^a_{d-(i-1)}$,
where $i=1,2,\ldots,d-3$. Thus the above prescription implies the 
identifications $j_d^a=j^{a+1}_d$, $j_1^a=j^{a-1}_d$, $j^a_{d-(i-1)}
=j^{a+i}_i$ among $j$ variables, while the glueing has to be accomplished by 
summing
over all free $m$ entries. 
The cyclic property of the joining implies that
the procedure is actually independent of the label $a$ chosen at the
beginning. Moreover, by requiring that the unique $\tilde{\sigma}^d$, which shares
$\sigma_{a}^{d-1}$ 
with $\sigma^d$, has indeed the same graph associated with its own $\sigma_{a}^{d-1}$, 
but different magnetic numbers with respect to the other one, 
we obtain the diagram shown in FIG.14 
\begin{figure}[htb]
\leavevmode
\hspace{2.5cm} \epsfbox{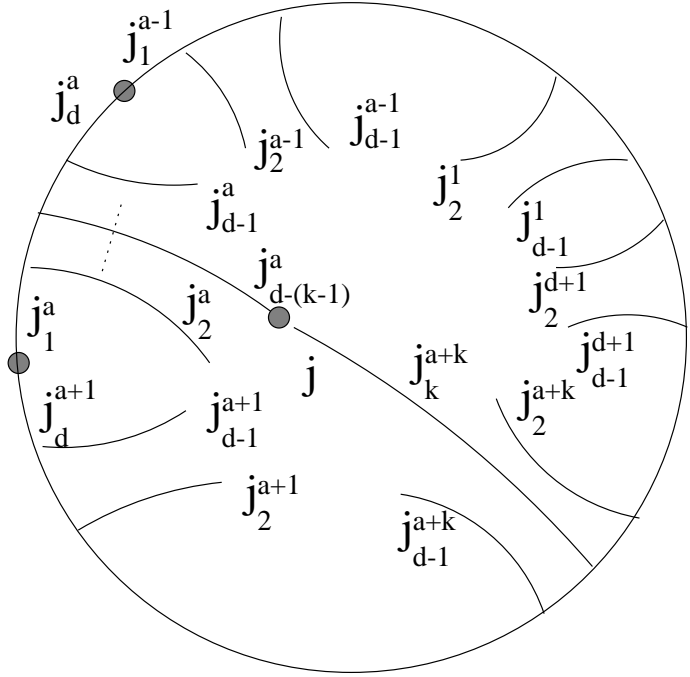}
\caption{ Diagram representing the 3nj symbols with its internal structure.}
\label{3nj}
\end{figure}
(which, by the above remarks, turns out to be
generic). 
A closer inspection of its combinatorial structure
shows that we get in fact a $\{3(d-2)(d+1)/2\}j$ symbol written in terms of (sums 
of)
$3jm$ symbols (see {\em e.g.} \cite{lituani}).\par
\noindent  Collecting all the previuos remarks we are now able to build up 
a state sum for $(T^d,\partial T^d)$ on the basis of the requirements listed 
below.\par
\begin{itemize}
\item 
For each $\sigma^d \in (T^d,\partial T^{d-1})$ we introduce:\par
\noindent {\em i)} an admissible set of colorings on each of its $(d-2)$-faces, 
namely
$j_1,j_2,$ $\ldots,$ $j_F$, where the value of the binomial coefficient
\begin{equation}
\left(\begin{array}{c}d+1 \\
2 \end{array}\right)\;\equiv \;F\nonumber
\end{equation}
\noindent gives the number of $(d-2)$ subsimplices of a $d$-simplex;\par
\noindent {\em ii)} other sets of $SU(2)$-colorings associated with
each of its $(d-1)$-faces and denoted collectively by $\{J_{i_1}\}$,
$\{J_{i_2}\}$, $\ldots$, $\{J_{i_{d+1}}\}$ (these sets are the counterparts of 
the five labels $J_a,\ldots,J_e$ used in (\ref{Jquasim})). Then we set
\begin{equation}
\sigma^d \longleftrightarrow \left\{\frac{3}{2}(d-2)(d+1)j\right\}_{\sigma^d}
\;\doteq\;\left[\{J_{i_1}\},\{J_{i_2}\},\ldots, \{J_{i_{d+1}}\}\right]_{\sigma^d} 
\label{dblock}
\end{equation}
\item
For each $\sigma^{d-1} \subset \partial T^d$, denoting as usual by $\{j',J'\}$
$\subset \{j,J\}$ the subsets of spin variables belonging to boundary components,
and labelling as $J'_1,$ $J'_2,$ $\ldots,$ $J'_C$ $(C=d-3)$ 
the variables associated 
with the internal legs of FIG.13, we have the explicit correspondence
\begin{eqnarray}
\lefteqn{  
\sigma^{d-1}\longleftrightarrow \sum_{{\cal M}}(-1)^{\sum_{C=1}^{d-3}{\cal M}_C}\;
\left( \begin{array}{ccc}
j'_1 & j'_2 & J'_1 \\
m_1 & m_2 & {\cal M}_1
\end{array}\right)\cdot}\hspace{.5cm}\nonumber\\
& & \cdot 
\left( \begin{array}{ccc}
J'_1 & j'_3 & J'_2 \\
-{\cal M}_1 & m_3 & {\cal M}_2
\end{array}\right)\cdot \cdot \cdot
\left( \begin{array}{ccc}
J'_{d-4} & j'_{d-2} & J'_{d-3} \\
-{\cal M}_{d-3} & m_{d-2} & {\cal M}_{d-3}
\end{array}\right)\cdot\nonumber\\
& & \cdot \left( \begin{array}{ccc}
J'_{d-3} & j'_{d-1} & j'_d \\
-{\cal M}_{d-3} & m_{d-1} & m_d
\end{array}\right).
\label{faceblock}
\end{eqnarray}
\noindent Here we agree that each $m$ variable is associated with the 
corresponding $j'$
on the upper row,
while an ${\cal M}$ entry is the magnetic number of the upper $J'$, with the usual 
ranges of
variations in both cases.
\end{itemize}
\noindent Then we can define the following state sum\par
\begin{eqnarray}
\lefteqn{Z[(T^d(j_{\sigma^{d-2}},J_{\sigma^{d-1}}), 
\partial T^d(j'_{\sigma^{d-2}},J'_{\sigma^{d-1}}
;m,{\cal M}))] \rightarrow (M^d,\partial M^d); L] =}\hspace{.2cm}\nonumber \\
& & =w_L^{(-1)^d \Xi}\,\prod_{all\,\sigma^{d-2}} (-1)^{2j_{\sigma^{d-2}}}
(2j_{\sigma^{d-2}}+1)
\prod_{all\,\sigma^{d-1}}
\left( \prod_{C=1}^{d-3}(-1)^{2J_C}(2J_C+1)\right)_{\sigma^{d-1}}
\cdot\nonumber\\
& & \cdot \prod_{all\, \sigma^d} 
\left\{\frac{3}{2}(d-2)(d+1)j\right\}_{\sigma^d}(J,J')
\prod_{\sigma^{d-1} \in \partial T^d}\sum_{\cal M}(-1)^{\sum m/2+ 
\sum {\cal M}}\cdot\nonumber\\
& & \cdot\left( \begin{array}{ccc}
j'_1 & j'_2 & J'_1 \\
m_1 & m_2 & {\cal M}_1
\end{array}\right)\cdot \cdot \cdot\left( \begin{array}{ccc}
J'_{d-3} & j'_{d-1} & j'_d \\
-{\cal M}_{d-3} & m_{d-1} & m_d
\end{array}\right),
\label{boundsum}
\end{eqnarray}
\noindent where  $\Xi =2(N_0-N_1+\ldots+(-1)^{d-3}N_{d-3})$, 
$(N_0,N_1,N_2,\ldots)$
being the total number of $(0,1,2,\ldots)$-dimensional simplices.
Notice also that some of the recoupling coefficients associated with
$d$-simplices may depend also on $J'$
variables, if they have components in $\partial T^d$.\par
\noindent The limiting procedure for handling all state sums (\ref{boundsum}) 
for the given $(M^d,\partial M^d)$ can be defined as\par
\begin{eqnarray}
\lefteqn{Z[(M^d,\partial M^d)]=}\hspace{-1cm}\nonumber\\
& & =\lim_{L\rightarrow \infty}\,\sum_{
\left\{\begin{array}{c}
T^d,\partial T^d\\
j,J\leq L
\end{array}\right\}}
Z[(T^d(j,J), \partial T^d(j',J';m,{\cal M})) \rightarrow
(M^d,\partial M^d);L],
\label{boundinv}
\end{eqnarray}
\noindent where the ranges of the magnetic quantum numbers are $|m|\leq j',
|{\cal M}|\leq J'$ in integer steps, and where suitable shorthand notations have 
been employed.\par
As anticipated before, we turn now to the basic question of the equivalence
of $Z[(M^d,\partial M^d)]$ in (\ref{boundinv}) under a suitable class of
topological operations. In the present case we are going to implement the
set of elementary inverse shellings $[n_{d-1} \rightarrow 
d-(n_{d-1})]^{{\bf d}}_{ish}$ in (\ref{dshel}), involving the attachment of
one $d$-simplex to $\partial T^d$ along $n_{d-1}=1,2,\ldots,d$
simplices of dimension $(d-1)$ glued together in suitable configurations. 
The complementary set
of the elementary shellings could be singled out simply by exchanging 
internal and external labellings in a consistent way. 
It should be clear from similar discussions on 
equivalence in {\em Section 3}, {\em Section 4} and {\em Appendix A}
that the explicit expressions of the identities associated with the moves
become more and more complicated as dimension grows. Thus, we limit ourselves
to the implementation of the moves through the diagrammatical method, 
which has been already used in {\em Section 4}. As a further remark, we 
note that glueing operations
performed on triangulations underlying $PL$-pairs of manifolds (as happens in our 
context)
involve joining of couples of $p$-dimensional simplices ($p=d,d-1$)
along their {\em unique} common $(p-1)$-dimensional face. More precisely, also
the joining of two $(d-1)$-dimensional simplices in $\partial T^d$ has to
fullfill this rule, since $\partial T^d$ is a manifold (indeed, this would be true
for pseudomanifols as well).\par
\noindent Before dealing with the full $d$-dimensional case, let us illustrate 
the case of elementary inverse shellings in $d=5$. Recall that a $5$-simplex
has six $4$-dimensional simplices in its boundary and that in the present case 
$n_{d-1}$ in (\ref{dshel}) may run over $1,2,3,4,5$. 
Consider first the inverse shelling  $[5\rightarrow 1]^{{\bf 5}}_{ish}$, the 
action of which is displayed in the diagram of FIG.15 
\begin{figure}[htb]
\hspace{1.5cm} \epsfbox{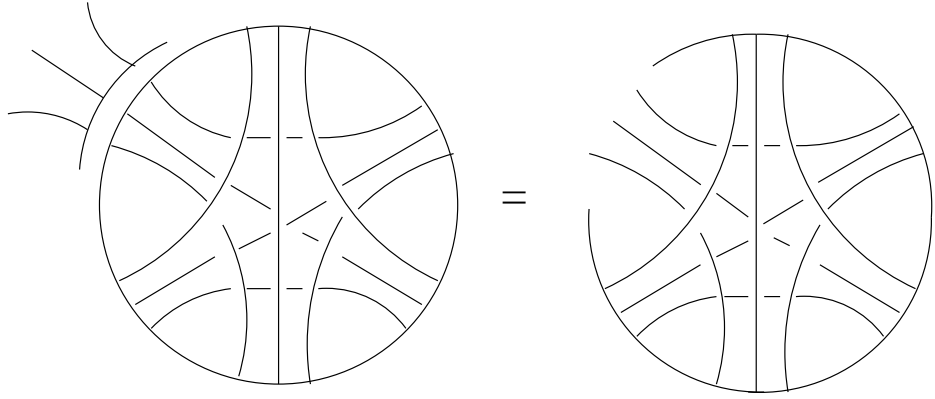}
\caption{ Shelling of $[5\rightarrow 1]_{ish}^{5}$ type.}
\label{51shel}
\end{figure}
(where we 
have made use 
of the diagrams shown in FIG.10 of {\em Section 3}). 
This operation amounts to glue a $5$-simplex to $\partial T^5$
along five $4$-simplices (joined among them along $3$-dimensional
faces). The resulting configuration in the modified $\partial T'^5$ gives rise to 
an unique new (open) $4$-simplex, so that no new $(5-r)$-simplices ($r\geq 3$) 
appear.
Thus the state sum (\ref{boundsum}) does not acquire any $w^{-2}_L$ 
factor and is
manifestely invariant under such a move.\par
\noindent The inverse shelling $[4\rightarrow 2]^{{\bf 5}}_{ish}$ consists in 
the attachment of a $5$-simplex along four $4$-simplices in $\partial T^5$. In
the new $\partial T'^5$ there appear two $4$-simplices joined along a common
$3$-simplex (for what we said before), and thus also in this case we do not 
introduce
new $(5-r)$-dimensional subsimplices ($r\geq 3$) in the state sum and no
additional $w^{-2}_L$ factor arises. The diagrammatic proof is given in
FIG.16,
\begin{figure}[htb]
\hspace{1.5cm} \epsfbox{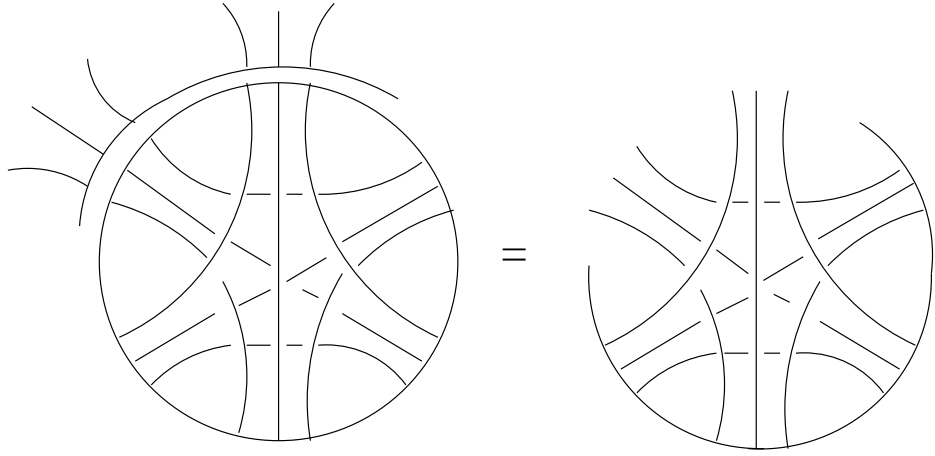}
\caption{ Shelling of $[4\rightarrow 2]^{5}_{ish}$ type.}
\label{42shel}
\end{figure}
where we have taken into account FIG.10 again.\par
\noindent The $[3\rightarrow 3]^{{\bf 5}}_{ish}$ move amounts to the glueing of
a $5$-simplex along three $4$-simplices lying in $\partial T^5$. 
Owing to the general remark that in a $p$-simplex $\subset \sigma^d$,
exactly three $(p-2)$-dimensional subsimplices incide over a $(d-3)$-dimensional
subsimplice, we see that such an inverse shelling generates just one new
triangle in $\partial T'^5$ (and no additional $(5-r)$-simplices with $r\geq 4$),
associated with a $w^{-2}_L$ factor in the state sum. The action of this move
is reproduced in FIG.17, 
\begin{figure}
\leavevmode
\hspace{3cm} \epsfbox{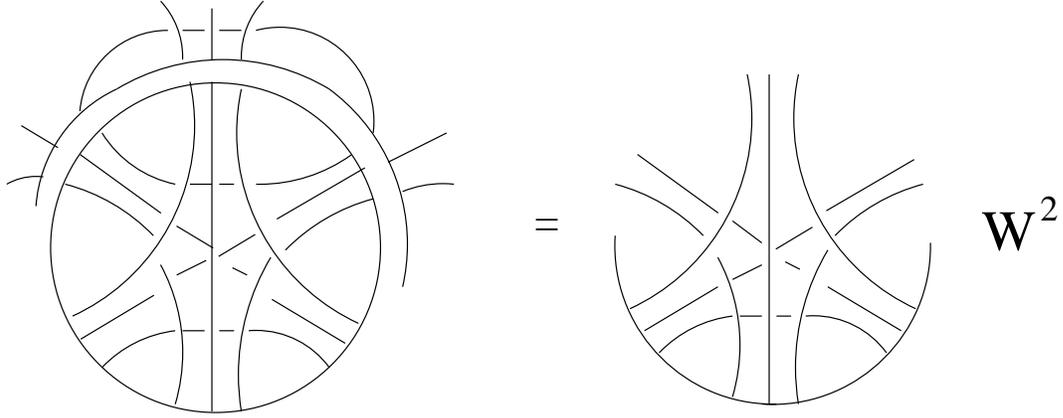}
\caption{ Shelling of $[3\rightarrow 3]^{5}_{ish}$ type.}
\label{33shel}
\end{figure}
where we can see the loop bringing a $w^{2}_L$
factor  which cancels the contribution coming from the
new triangle.\par
\noindent The move $[2\rightarrow 4]^{{\bf 5}}_{ish}$ represents the attachment of
a $5$-simplex to $\partial T^5$ along four $4$-dimensional simplices. Recalling
the expressions giving the number of subsimplices of a $p$-simplex (see below),
we may see that the configuration of two $4$-simplices, glued along their common
$3$-dimensional face, identifies six vertices, fourteen edges, sixteen triangles
and nine tetrahedra; thus this type of inverse shelling generates in $\partial 
T'^5$
one new edge ($(d-4)$-simplex) and four triangles ($(d-3)$-simplices), giving rise
to $w^{-6}_L$ factor in the state sum. The above action in depicted in the diagram 
of FIG.18, 
\begin{figure}[htb]
\leavevmode
\hspace{3cm} \epsfbox{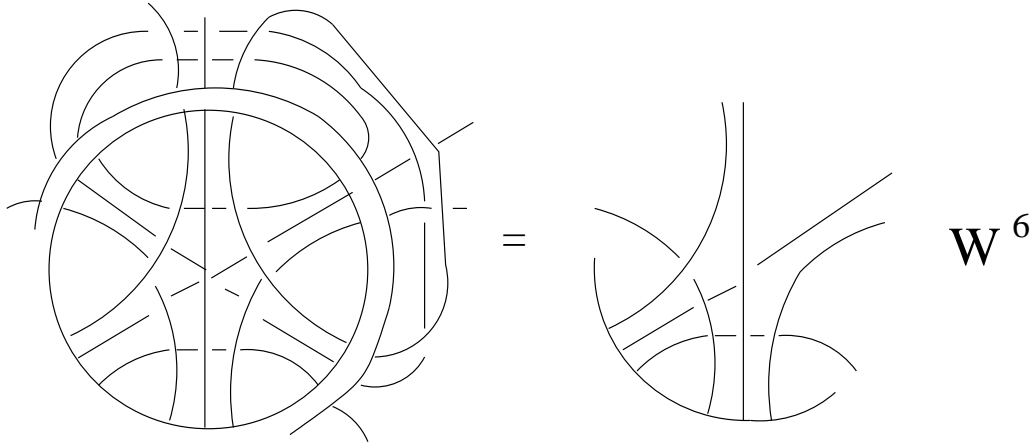}
\caption{ Shelling of $[2\rightarrow 4]^{5}_{ish}$ type.}
\label{24shel}
\end{figure}
where we see three loops, the contributions of which cancel the 
above
extra factor.\par
\noindent The last type of inverse shelling that we deal with is
$[1\rightarrow 5]^{{\bf 5}}_{ish}$, representing the glueing of a $5$-simplex 
along
one $4$-simplex. By counting in an appropriate way the subsimplices of the new
configuration in $\partial T'^5$, we see that there appear ten triangles 
($(d-3)$-simplices), five edges ($(d-4)$-simplices) and one vertex 
($(d-5)$-simplex),
which generate an overall $w^{-12}_L$ factor. Looking now at FIG.19, 
\begin{figure}[htb]
\leavevmode
\hspace{3.5cm} \epsfbox{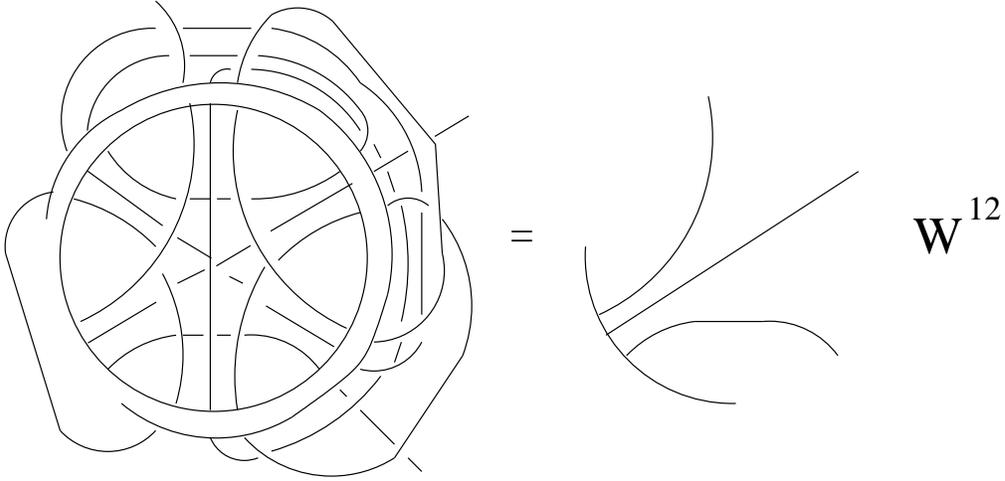}
\caption{ Shelling of $[1\rightarrow 5]^{5}_{ish}$ type.}
\label{15shel}
\end{figure}
such extra 
factor
turns out to be exactly cancelled by the contributions arising from the loops.
This completes the proof of the invariance of (\ref{boundinv}) 
under elementary boundary operations in the $5$-dimensional case.\par

Coming to the general $d$-dimensional case, we slightly change our previous 
notation, 
namely $[n_{d-1} \rightarrow d-(n_{d-1})]^{{\bf d}}_{ish}$
($n_{d-1}=1,2,\ldots,d$), by
parametrizing the moves in terms of $d$ according to\par
\begin{equation}
[(d-k) \rightarrow (k+1)]^{{\bf d}}_{ish}\:\:,k=0,1,2,\ldots,(d-1),
\label{dinsh}
\end{equation} 
\noindent which of course turns out to be consistent with the previous one.
Notice also that in what follows we shall make use of the diagrammatic 
relations shown in FIG.9 and FIG.10 whenever it is necessary.\par 
\noindent Consider first $[d\rightarrow 1]^{{\bf d}}_{ish}$, representing the 
glueing of
a $d$-simplex to $\partial T^d$ along $d$ $(d-1)$-dimensional simplices (the 
corresponding diagram is given in FIG.20). 
\begin{figure}[htb]
\leavevmode
\hspace{5.5cm} \epsfbox{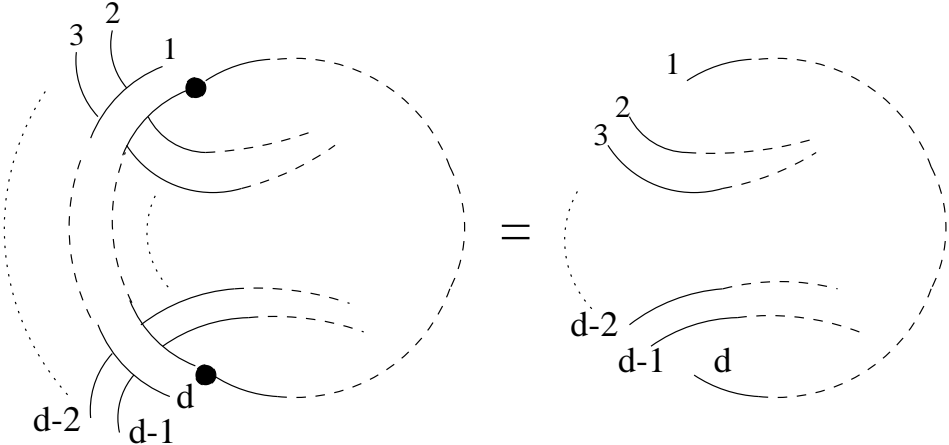}
\caption{ Shelling of $[d \rightarrow 1]^{{\bf d}}_{ish}$ type.}
\label{n1shel}
\end{figure}
Since we do not generate any 
$(d-r)$-simplex 
($r \geq 3$) in the new configuration $\partial T'^d$, no additional $w^{-2}_L$ 
factors 
enter the state sum (\ref{boundsum}) .\par
\noindent The $[(d-1)\rightarrow 2]^{{\bf d}}_{ish}$ move consists in the 
attachment
of a $d$-simplex to $\partial T^d$ along $(d-1)$ simplices of dimension $(d-1)$.
Also the action of this inverse shellings does not give any $(d-r)$-simplex 
($r\geq 3$) 
in the new $\partial T'^d$, and its graphical counterpart is shown in 
FIG.21.\par
\begin{figure}[htb]
\leavevmode
\hspace{2.5cm} \epsfbox{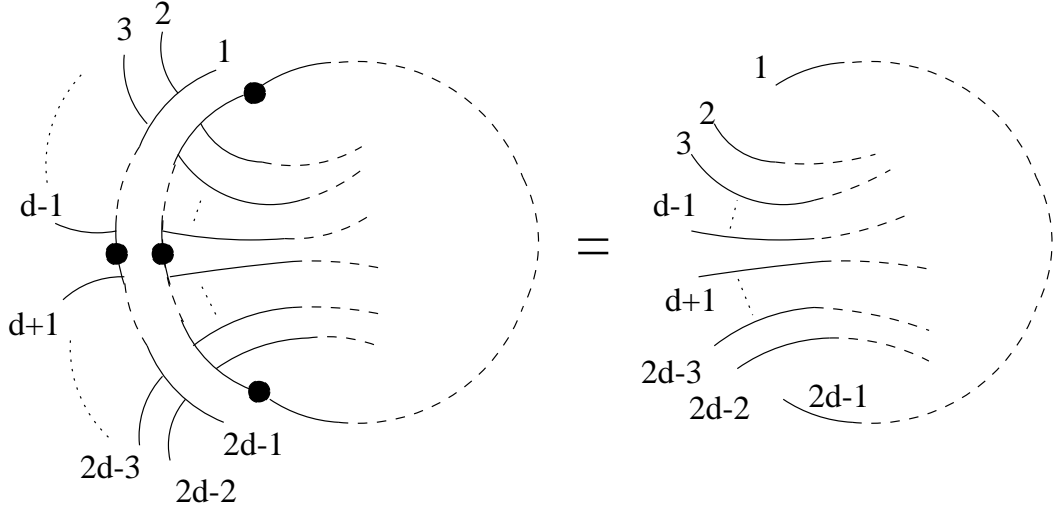}
\caption{ Shelling of $[d-1 \rightarrow 2]^{{\bf d}}_{ish}$  type.}
\label{n-12shel}
\end{figure}
\noindent Coming to $[(d-2)\rightarrow 3]^{{\bf d}}_{ish}$, we see that it 
represents the glueing of a $d$-simplex along $(d-2)$ simplices of dimension 
$(d-1)$ lying in
$\partial T^d$ and its diagram is given in FIG.22. 
\begin{figure}[htb]
\leavevmode
\hspace{2.5cm} \epsfbox{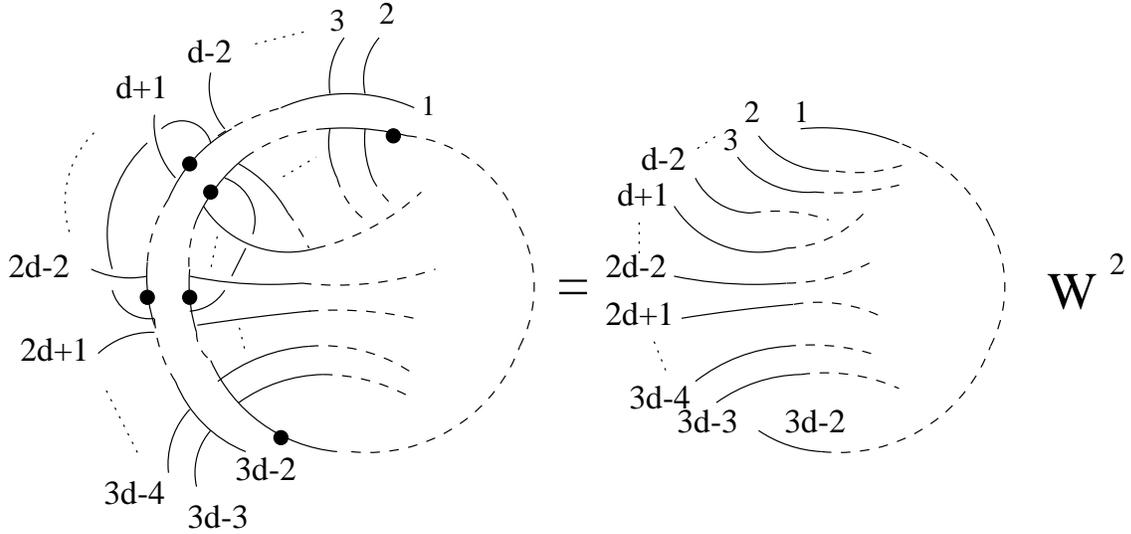}
\caption{ Shelling of $[d-2 \rightarrow 3]^{{\bf d}}_{ish}$  type.}
\label{n-23shel}
\end{figure}
In the new boundary
triangulation one $(d-3)$-simplex (and no other simplices) appears. Thus the
$w^{-2}_L$ factor which comes out is exactly cancelled by the contribution of the 
loop.\par
\noindent An algorithmic setting can now be easily established. For,
the $k$-th inverse shelling, namely $[(d-k) \rightarrow (k+1)]^{{\bf d}}_{ish}$,
represents the attachment of a new $d$-simplex along $(d-k)$
simplices of dimension $(d-1)$ in $\partial T^d$ which generates
in $\partial T'^d$ several kinds of new boundary components.
The number of such new components is evaluated by using suitable binomial 
coefficients,
according to the list given below\par
\begin{equation}
\left(\begin{array}{c}
k+1\\3\end{array}\right)
\,\sigma^{d-3};\:
\left(\begin{array}{c}
k+1\\4\end{array}\right)
\,\sigma^{d-4};\:
\ldots;\:
\left(\begin{array}{c}
k+1\\k+1\end{array}\right)
\,\sigma^{d-(k+1)},
\label{binomial}
\end{equation}
\noindent while for each $k$ the following number of additional $w^{-2}_L$ 
factors are generated \par
\begin{equation}
w_L^{2\sum_{i=3}^{k+1}\left(\begin{array}{c}
k+1\\i\end{array}\right)(-1)^{i+1}}\;=\;
w_L^{-2\frac{k-1}{2}k}.
\label{pesi}
\end{equation}
\noindent The action of the $k$-th inverse shelling is depicted in the 
diagram of FIG.23, 
\begin{figure}[htb]
\leavevmode
\hspace{2.5cm} \epsfbox{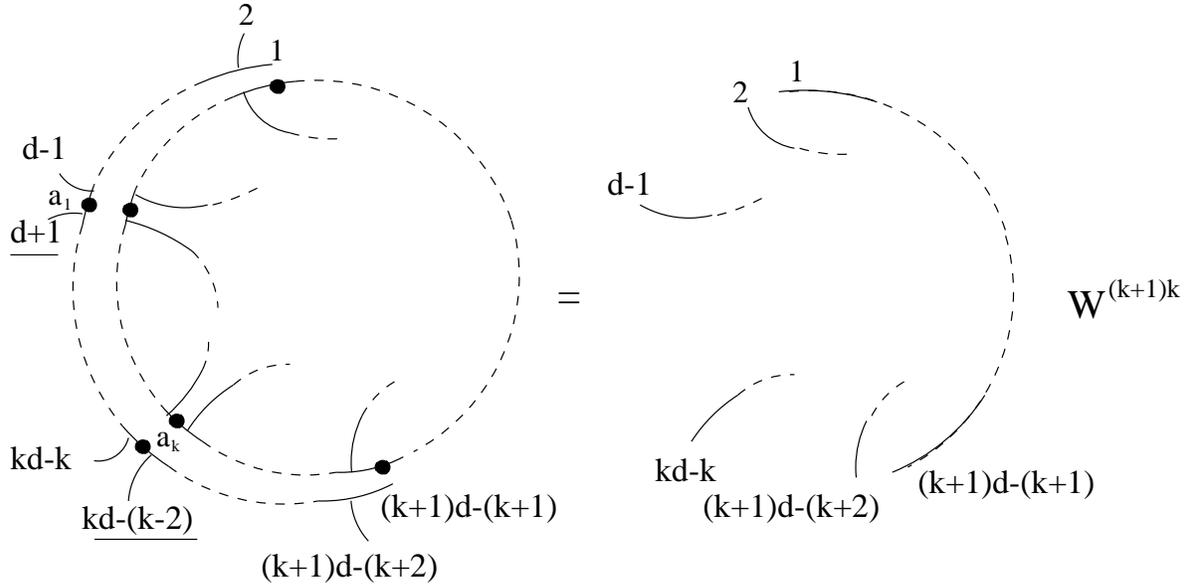}
\caption{ Shelling of $[d-k \rightarrow k+1]^{{\bf d}}_{ish}$  type.}
\label{kkshel}
\end{figure}
where
the loops which cancel the weights given in (\ref{pesi}) appear. 
More precisely, when we glue together $(k+1)$ $(d-1)$-simplices in the more symmetric way, 
we have to perform  $(\sum_{l=1}^k l)$ identifications among their faces 
(and that is exactly the number of
summations over $j$ variables in the state sum).
However, $k$ of the above identifications do not bring loops (as can be
inferred from
the structure of the $\{3(d-2)(d+1)/2\}j$ symbol) and thus the remaining
($\sum_{l=1}^{k-1}l$) glueing operations induce exacly  the factors given in (\ref{pesi}).
This remark completes the proof of the
equivalence of (\ref{boundsum}) under the entire set of boundary
elementary operations. Thus, by Pachner's result found in \cite{pachner90},
the expression given in (\ref{boundinv}) is formally an invariant of the
$PL$-structure of $(M^d,\partial M^d)$. Its regularized counterpart,
$Z[(M^d,\partial M^d)](q)$ can be written explicitly according to the prescription
given in {\em Appendix B}.\par


\section{Invariants of closed $M^d$ from colorings of \\
$(d-2)$-simplices}

In order to complete the program outlined in the introduction, in the present 
section we show how it it possible to generate from  the extended invariant
defined in {\em Section 5} a state sum model for a closed $PL$-manifold 
$M^d$. Obviously the form of the new state sum will be inferred from
the extended one (given in (\ref{boundsum})) simply by ignoring the contributions from
boundary ({\em cfr.} the $3$ and $4$-dimensional cases). 
Thus the main point of the present section will consist in giving the proof that
(\ref{boundinv}), and consequently $Z[M^d]\doteq Z[(M^d,\partial M^d)]|_{\partial 
M^d
=\emptyset}$, are invariant under $d$-dimensional bistellar moves performed either
in the interior (bulk) of a triangulation $(T^d,\partial T^d)$
or in a closed $T^d$, respectively.\par
\noindent The expression of the state sum for a triangulation $T^d$, colored 
according to the prescription of the previous section and with the same
notations, reads\par
\begin{eqnarray}
\lefteqn{Z[T^d(j_{\sigma^{d-2}},J_{\sigma^{d-1}}) 
\rightarrow M^d; L] =}\hspace{.5cm}\nonumber \\
& & =w_L^{(-1)^d \Xi}\,\prod_{all\,\sigma^{d-2}} (-1)^{2j_{\sigma^{d-2}}}
(2j_{\sigma^{d-2}}+1)
\prod_{all\,\sigma^{d-1}}
\left( \prod_{C=1}^{d-3}(-1)^{2J_C}(2J_C+1)\right)_{\sigma^{d-1}}
\cdot\nonumber\\
& & \cdot \prod_{all\, \sigma^d} 
\left\{\frac{3}{2}(d-2)(d+1)j\right\}_{\sigma^d}(J),
\label{closesum}
\end{eqnarray}
\noindent while the limit taken on all admissible colored triangulations
of a given $PL$-manifold $M^d$ is formally written as follows\par
\begin{equation}
Z[M^d]\,=\,\lim_{L\rightarrow \infty}\,\sum_{
\left\{\begin{array}{c}
T^d\\
j,J\leq L
\end{array}\right\}}
Z[T^d(j,J) \rightarrow M^d;L].
\label{closeinv}
\end{equation}
The issue of the equivalence of (\ref{closeinv})
under the suitable set of topological operations can be addressed
by exploiting some results from $PL$-topology (recall also the definitions given
at the beginning of the introduction) and on applying them to
the extended state sum $Z^d[(T^d,\partial T^d)\rightarrow \ldots]$
given in (\ref{boundsum}).\par 
\noindent Let us start by considering the simplicial complex made up by
one$d$-simplex $\sigma^d$ together with all its subsimplices. From the topological
point of view we get in fact what is called a (standard) simplicial $PL$-ball, 
and we denote it by ${\cal B}^d(\sigma)$ (we omit the dimensionality 
of simplices whenever it is clear from the context). The boundary of such a ball,
$\partial {\cal B}^d(\sigma)$, is obviously homeomorphic to the 
$(d-1)$-dimensional
sphere and in particular it is a simplicial $PL$-sphere containing the $(d+1)$ 
faces
of dimension $(d-1)$ of the original ${\cal B}^d(\sigma)$. 
Notice however that a simplicial $(d-1)$-sphere can be defined by its own by
joining in a suitable way some  $(d-1)$-dimensional simplices $\subset {\Bbb 
R}^d$.
The minimum number of $(d-1)$-simplices necessary to get a $PL$-sphere is
just $(d+1)$: the resulting simplicial sphere will be denoted by 
${\cal S}^{d-1}(\sigma_1\cup \sigma_2\cup \ldots \cup \sigma_{d+1})$. If we 
consider again
the $PL$-ball ${\cal B}^d(\sigma)$, we would get 
$\partial {\cal B}^d(\sigma)$ $\cong_{PL}$ ${\cal S}^{d-1}(\sigma_1\cup 
\sigma_2\cup \ldots \sigma_{d+1})$, where $\cong_{PL}$ stands for a 
$PL$-homeomorphism.\par
\noindent Turning now to the structure of the extended state sum for a 
$(T^d,\partial T^d)$, we can reconsider its topological content as follows. 
Indeed,
we see that the contribution of the configuration ${\cal S}^{d-1}(\sigma_1\cup 
\sigma_2\cup \ldots \cup \sigma_{d+1})$ to $Z^d$ amounts exactly to one 
$\{3(d-2)(d+1)/2\}j$ symbol,
namely it is the same that we would obtain by glueing a ${\cal B}^d(\sigma)$ to
${\cal S}^{d-1}(\sigma_1\cup \sigma_2\cup \ldots \cup \sigma_{d+1})$ along
$\partial {\cal B}^d(\sigma)$ with a $PL$-homeomorphism.
The reason why we stress this point relies on the fact that on this basis
we are able to set up the following step--by--step procedure: {\em i)} we extract 
first
some ${\cal B}^d(\sigma)$ from the bulk of $(T^d, \partial T^d)$, leaving an 
internal hole
bounded by the $PL$-sphere ${\cal S}^{d-1}(\sigma_1\cup \sigma_2\cup \ldots 
\sigma_{d+1})$;
{\em ii)} then we carry out elementary boundary operations on the $PL$-pair
$({\cal B}^d(\sigma)$ $,\partial {\cal B}^d(\sigma))$ bringing $\partial {\cal 
B}^d(\sigma)$ 
into ${\cal S}^{d-1}(\tau_1\cup \tau_2\cup \ldots \cup \tau_{d+1})$ (notice that 
in
doing that we do not alter the extended state sum, owing to its invariance under 
elementaty shellings); {\em iii)} finally, we glue the ball back into the original 
triangulation 
through a $PL$-homeomorphism 
${\cal S}^{d-1}$ $(\sigma_1\cup \sigma_2\cup \ldots \cup \sigma_{d+1})$ 
$\cong_{PL}$
${\cal S}^{d-1}$ $(\tau_1\cup \tau_2\cup \ldots \cup \tau_{d+1})$.\par
\noindent Such kinds of {\em cut and paste} represent nothing that
the implementation of the set of $d$-dimensional bistellar moves in the bulk of
each triangulation of $(M^d,\partial M^d)$ (and in the whole closed $M^d$).
To be precise, the entire set of moves will be obtained by cutting away not just a
standard $PL$-ball as before, but rather simplicial balls
made up of a suitable collection of more than one $d$-simplex.\par
\noindent As an explicit example of how the above procedure works, consider the 
$3$-dimensional case with the corresponding extended state sum
given in (\ref{clsum2}). Recall from (\ref{dbst}) that in this case we
are dealing with four bistellar moves, $[1\leftrightarrow 4]^{{\bf 3}}_{bst}$
and $[2\leftrightarrow 3]^{{\bf 3}}_{bst}$, where the arguments refer to the 
number
of $3$-simplices involved in the corresponding transformation. The explicit
implementation of some of these moves is  given below.\par
\begin{itemize}
\item $[1\rightarrow 4]^{{\bf 3}}_{bst}$. Since the initial configuration contains 
just one
$3$-simplex, we are just in the situation described above. Then we
extract the ball ${\cal B}^3(\sigma^3)$ (the boundary of which is 
${\cal S}^2$ $(\sigma^2_1\cup \sigma^2_2
\cup \sigma^2_3 \cup \sigma^2_4)$) and
perform on it the inverse shelling $[1\rightarrow 3]^{{\bf 3}}_{ish}$, where the
first triangle is chosen in an arbitrary way. Thus we get a configuration with
two $3$-simplices, namely the original $\sigma^3$ and a new $\tau^3_1$, glued along
the original triangle. The second operation is an inverse shelling
$[2\rightarrow 2]^{{\bf 3}}_{ish}$, where the two initial contiguous triangles
belong to $\sigma^3$ and to $\tau^3_1$, respectively. This move generates a third 
tetrahedron,
$\tau^3_2$, joined to the previous ones through a $2$-dimensional face. 
On this configutation we act now with
$[3\rightarrow 1]^{{\bf 3}}_{ish}$, where two of the three triangles of the
initial arrangment were generated in the two previous steps, respectively, while
the third one belongs to the original $\sigma^3$. Thus we get a fourth tetrahedron
$\tau^3_3$ which, together with the other ones, gives rise to the simplicial
ball  ${\cal B}^3(\sigma^3$ $\cup \tau^3_1 \cup$ $\tau^3_2\cup \tau^3_3)$,
the boundary of which is 
${\cal S}^2(\sigma^2\cup 
\tau^2_{(1)}
\cup \tau^2_{(2)} \cup \tau^2_{(3)})$, where the first entry is the initial 
triangle
chosen in $\sigma^3$, and the other entries are the new faces generated in the 
three
previous steps ($\tau^2_{(1)} \subset \tau^3_1; \ldots$). 
Now we glue back the modified simplicial ball into the triangulation through a
$PL$-homeomorphism between the original ${\cal S}^2$ $(\sigma^2_1\cup 
\sigma^2_2
\cup \sigma^2_3 \cup \sigma^2_4)$ and
${\cal S}^2(\sigma^2\cup 
\tau^2_{(1)}
\cup \tau^2_{(2)} \cup \tau^2_{(3)})$. The pictorial representation of the 
reconstruction
of this particular move is shown in FIG.24.\par
\begin{figure}[htb]
\leavevmode
\hspace{5cm} \epsfbox{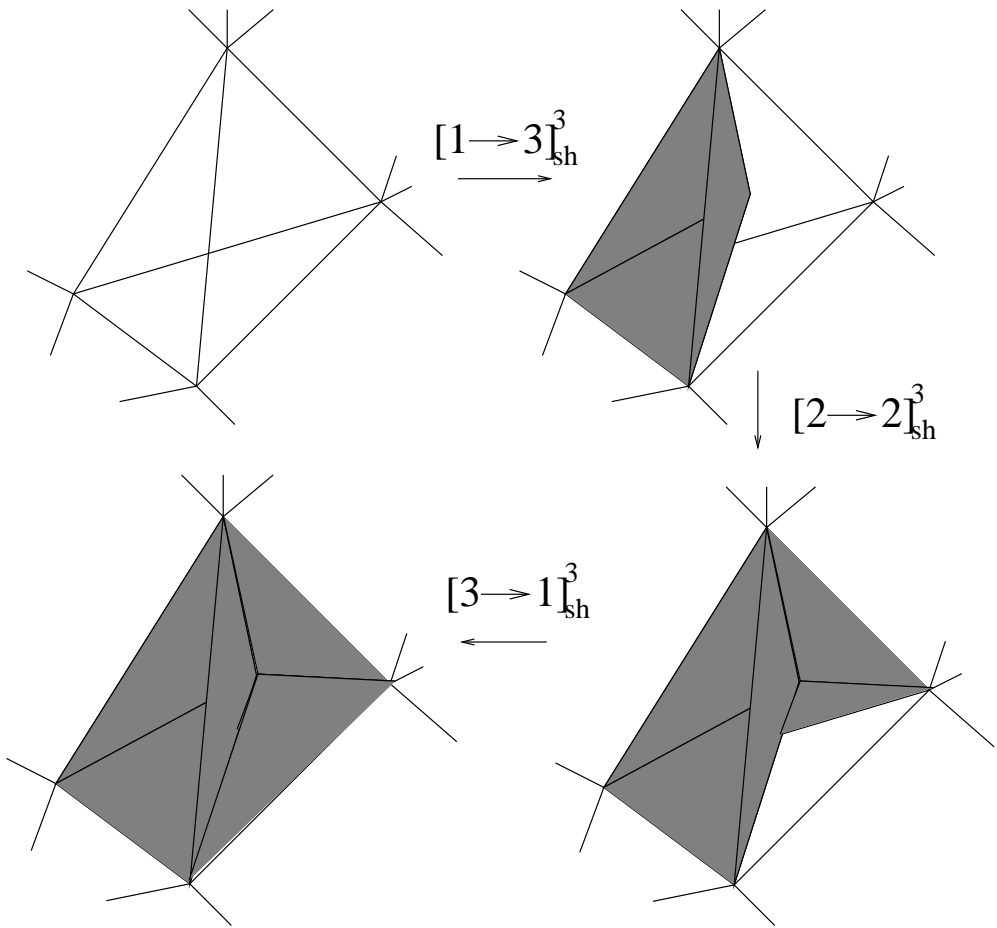}
\caption{ The move $[1 \rightarrow 4]^{\bf 3}_{bst}$ in terms of inverse
shellings performed on the removed PL-ball.}
\label{bi14}
\end{figure} 
\item $[2\rightarrow 3]^{{\bf 3}}_{bst}$. The configuration we start with is
a ball ${\cal B}^3(\sigma_1^3 \cup \sigma_2^3)$ having  ${\cal S}^2$ 
$(\sigma^2_1\cup 
\sigma^2_2
\cup \sigma^2_3 \cup \sigma^2_4 \cup \sigma^2_5 \cup \sigma^2_6 )$ as its 
boundary. 
After the extraction, we perform first the inverse shelling 
$[2\rightarrow 2]^{{\bf 3}}_{ish}$, where the two initial triangles are contiguous 
and
belong to $\sigma^3_1$ and to $\sigma^3_2$, respectively. Finally, we apply
$[3\rightarrow 1]^{{\bf 3}}_{ish}$, where the three initial triangles belong to
$\sigma^3_1$, $\sigma^3_2$ and to the component generated by the previous step,
respectively. It is not difficult to realize that the resulting $PL$-ball
has again six faces in its boundary $PL$-sphere, and thus we glue it back
along the boundary of the  original hole by means of a suitable 
$PL$-homeomorphism.
The sequence of operations we have performed is drawn in FIG.25.\par
\begin{figure}[htb]
\leavevmode
\hspace{2.5cm} \epsfbox{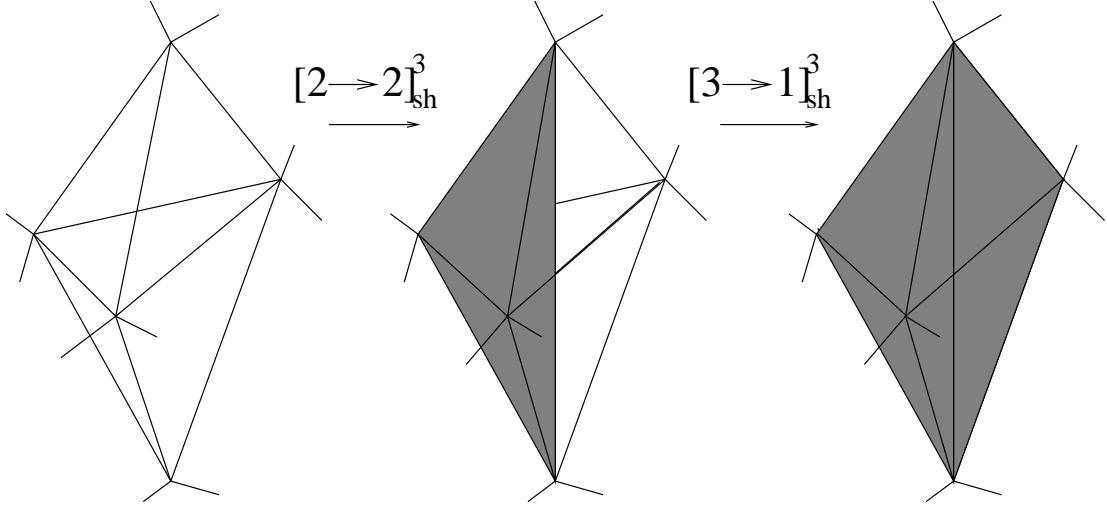}
\caption{  The move $[2 \rightarrow 3]^{\bf 3}_{bst}$ in terms of inverse
shellings performed on the removed PL-ball.}
\label{bi23}
\end{figure}
\end{itemize}
\noindent The remaining $3$-dimensional bistellar moves can be explicitly 
worked out following a similar procedure and by employing a definite 
sequence of inverse operations.\par
\noindent Also for what concerns the $4$-dimensional case we could describe step--by--step
the implementation of all transformation. However, since in that case
a pictorial counterpart is not so easy displayed, we turn to the general rule
in dimension $d$. On the basis of the characterization of $d$-dimensional inverse 
shellings
given in (\ref{dinsh}) we can reparametrize also the set of all $d$-bistellar 
moves 
according to\par
\begin{equation}
[k\rightarrow (d-k+1)]^{{\bf d}}_{bst}\;\;;k\,=\,1,2,\ldots,(d-1),
\label{kbst}
\end{equation}
\noindent paying attention to the fact that here $k$ enumerates $d$-simplices,
while in (\ref{dinsh}) it was referred to $(d-1)$-simplices in a given
$\sigma^d$. With this reformulation, we see that the generic move  
$[k\rightarrow (d-k+1)]^{{\bf d}}_{bst}$ can be obtained with the {\em cut and 
paste}
procedure following the steps:\par
\begin{itemize}
\item remove from the bulk of $(T^d,\partial T^d)$ a $PL$-ball ${\cal 
B}^d(\sigma_1 \cup$ $\sigma_2 \cup \ldots \cup \sigma_k)$,
where $k=1,2,\ldots,(d-1)$ ;\par
\item implement on the ball the sequence of inverse shellings   
$[k\rightarrow d-k+1]^{{\bf d}}_{ish}$, $[k+1\rightarrow d-k]^{{\bf d}}_{ish}$, 
$\ldots$, $[d+1\rightarrow 1]^{{\bf d}}_{ish}$, by choosing in the initial 
configuration
one $(d-1)$-simplex in each of the components of the original ball and
by involving in the subsequent moves one $(d-1)$-simplex for each of the components 
generated
before (including also the initial one);\par 
\item glue back the resulting modified simplicial ball along the hole left 
in $(T^d,\partial T^d)$
by using a $PL$--homeomorphism.\par
\end{itemize}
Having shown that the extended state sum is equivalent under (a finite set of)
bistellar moves performed in the bulk for any dimension $d$, 
we can conclude that the expression of $Z[M^d]$ in (\ref{closeinv}) share the
same property. Thus, owing to \cite{pachner87}, it is also an invariant
of the $PL$-stucture. For what concern the $q$-deformed $Z[M^d](q)$, we refer
as usual to the notations and definitions collected in {\em Appendix B}.

\section*{Conclusions}

Coming back to the array of $PL$-invariants displayed in the introduction, we would
hopefully complete it with a third hierarchy, namely
\vskip 0.5 cm
\begin{center}
$\begin{array}{cccccccc}
dimension: & 2 & 3 & 4 & \ldots \\
  &  &  &  &  &  \\
  & Z^2_{\chi} & Z^3_{PR} &  W^4    \\
  &  &  &  &  &  \\
  &  & Z^3_{\chi} & Z^4_{CY} &  \ldots  \\
  &  &  &  &  \ldots  \\
\end{array}$
\end{center}
\vskip 0.5 cm
As we learnt from the results found in the previous sections, the key tool to
build up the first entry of the new hierarchy, namely $W^4$, consists in rewriting
$Z^3_{PR}$ in terms of double symbols (to be associated with each tetrahedron).
We are currently investigating such a possibility, which could become concrete
by exploiting the Regge symmetries of the $6j$ symbol (see \cite{simmetrie}).
The next step will consist in picking up one of the sub--symbols of the double
symbol, collecting five of them in a suitable way, and associating the resulting 
expression with the fundamental $4$-dimensional block. Finally, the issue of the
equivalence of the resulting state sum under moves should be improved, together with
the identification of the {\em continuous} counterpart of the new $PL$-invariant.
Indeed, such a $W^4$ could be non--trivial since it would be easily extended to an
invariant for a pair $(M^4,\partial M^4)$ having $Z^3_{PR}$ on its boundary manifold.\par
As a last remark, we would like to spend some words on the procedure described in 
{\em Section 6}, where we established that the 
equivalence of the closed state sum under 
bistellar moves follows from the equivalence of the extended state sum under
shellings. It is worthwhile to notice that the 
possibility of carrying out topological operations 
on removed $PL$-balls relies on a theorem given in \cite{plballs} 
which states that {\em Every $d$-dimensional $PL$-ball is shellable}. On the
other hand, it is clear that the glueings of the modified balls into the
bulk of the original triangulation are achieved through $PL$-homeomorphisms which are
consistent with respect to  $SU(2)$-labellings but, generally speaking, are 
not at all isometric mappings (the natural metric being an Euclidean $PL$-metric induced
on the underlying polyhedron). Thus, if we were interested in finding a state sum
for a pair $(M^d,\partial M^d)$ in which the edge lengths of the simplices in the 
bulk are dynamical variables (whereas for istance we would keep on 
requiring invariance on the boundary manifold) then we should accordingly modify the whole approach.
We are currently addressing such issues in connection with the search for discretized
models of Euclidean quantum gravity in dimension four.

\vfill
\newpage
\section*{Appendix A}
We give here explicitly the identities corresponding to the four types of {\em
elementary shellings} in $d=4$ which are employed to show the 
invariance of $Z[(M^4, \partial M^4)]$ in {\em Section 3}. Recall from
(\ref{boutresim}) and (\ref{quadrisum}) that there we made use of primed
spin variables $\{j',J'\}$ in order to label those components which lie in
$\partial T^4$ in some configuration, while plain $\{j,J\}$ denoted 
components in $int (T^4)$. Since here almost all variables are indeed in
$\partial T^4$, we agree to change our previous notation according to\par
\vskip 0.5cm
$j'_1,j'_2,\ldots,j'_{10}\,\longrightarrow \,j_1,j_2,\ldots,j_{10}$,\par

$J'_a (\subset \partial T^4)\,\longrightarrow\,J_a$,\par

$J_a (\subset int(T^4))\,\longrightarrow\,{\bf J_a}$.\par
\vskip 0.5cm
\noindent For the corresponding $m$ variables we keep on denoting them by plain
$m_1,m_2,\ldots$, $m_{10}$; $m_a,\ldots,m_e$, since no ambiguity can arise.
We set also $w^2_j \equiv (2j+1)$, and the summation labels and
arguments are shortened as far as possible.\par
\vskip 3cm
\noindent ${\bf [1 \rightarrow 4]^4_{sh}}$ (see FIG.5)\par
\begin{eqnarray*}
\lefteqn{\sum_{J_a,m_a}w^2_{J_a} (-1)^{m_a}
\left( \begin{array}{ccc}
j_1 & j_2 & J_a \\
m_1 & m_2 & m_a
\end{array}\right)
\left( \begin{array}{ccc}
j_3 & j_4 & J_a \\
m_3 & m_4 & -m_a
\end{array}\right)
[J_a,{\bf J_b},{\bf J_c},{\bf J_d},{\bf J_e}]
 =}\hspace{.2cm}\nonumber \\
& & \cdot \sum_{m_i}(-1)^{\sum_{i}m_i} \sum_{m_A}(-1)^{\sum_{A}m_A}
\left( \begin{array}{ccc}
j_5 & j_6 & J_b \\
m_5 & m_6 & m_b
\end{array}\right)
\left( \begin{array}{ccc}
j_3 & j_7 & J_b \\
-m_3 & m_7 & -m_b
\end{array}\right)\cdot \\
& & \cdot \left( \begin{array}{ccc}
j_5 & j_8 & J_c \\
-m_5 & m_8 & m_c
\end{array}\right)
\left( \begin{array}{ccc}
j_1 & j_9 & J_c \\
-m_1 & m_9 & -m_c
\end{array}\right)
\left( \begin{array}{ccc}
j_6 & j_{10} & J_d \\
-m_6 & m_{10} & m_d
\end{array}\right)\cdot \\
& & \cdot \left( \begin{array}{ccc}
j_2 & j_8 & J_d \\
-m_2 & -m_8 & -m_d
\end{array}\right)
\left( \begin{array}{ccc}
j_7 & j_{10} & J_e \\
-m_7 & -m_{10} & m_e
\end{array}\right)
\left( \begin{array}{ccc}
j_4 & j_9 & J_e \\
-m_4 & -m_9 & -m_e
\end{array}\right)
\end{eqnarray*}
\vskip 0.5cm

$\{m_i\}=(m_5,m_6,m_7,m_8,m_9,m_{10})$

$\{m_A\}=(m_b,m_c,m_d,m_e)$.

\vskip 3cm
\noindent ${\bf [2 \rightarrow 3]^4_{sh}}$ (see FIG.6)\par
\begin{eqnarray*}
\lefteqn{\sum_{j_3,m_3}w^2_{j_3}(-1)^{m_3}
\sum_{J_A,m_A}(\prod_{A}w^2_{J_A}) (-1)^{\sum_{A}m_A}
\left( \begin{array}{ccc}
j_1 & j_2 & J_a \\
m_1 & m_2 & m_a
\end{array}\right)
\left( \begin{array}{ccc}
j_3 & j_4 & J_a \\
m_3 & m_4 & -m_a
\end{array}\right)
\cdot}\hspace{.2cm}\nonumber \\
& & \cdot \left( \begin{array}{ccc}
j_5 & j_6 & J_b \\
m_5 & m_6 & m_b
\end{array}\right)
\left( \begin{array}{ccc}
j_3 & j_7 & J_b \\
-m_3 & m_7 & -m_b
\end{array}\right)
[J_a,J_b,{\bf J_c},{\bf J_d},{\bf J_e}]=\\
& & =\sum_{m_i}(-1)^{\sum_{i}m_i} \sum_{m_A}(-1)^{\sum_{A}m_A}
\left( \begin{array}{ccc}
j_5 & j_8 & J_c \\
-m_5 & m_8 & m_c
\end{array}\right)
\left( \begin{array}{ccc}
j_1 & j_9 & J_c \\
-m_1 & m_9 & -m_c
\end{array}\right)\cdot\\
& & \cdot\left( \begin{array}{ccc}
j_6 & j_{10} & J_d \\
-m_6 & m_{10} & m_d
\end{array}\right)
\left( \begin{array}{ccc}
j_2& j_8 & J_d \\
-m_2 & -m_8 & -m_d
\end{array}\right)
\left( \begin{array}{ccc}
j_7 & j_{10} & J_e \\
-m_7 & -m_{10} & m_e
\end{array}\right)\cdot\\
& & \cdot \left( \begin{array}{ccc}
j_4 & j_9 & J_e \\
-m_4 & -m_9 & -m_e
\end{array}\right)
\end{eqnarray*}

$\{J_A\}=(J_a,J_b)$

$\{m_A\}=(m_a,m_b)$

$\{m_i\}=(m_8,m_9,m_{10})$

$\{m_B\}=(m_c,m_d,m_e)$.

\vskip 1cm

\noindent ${\bf [3 \rightarrow 2]^4_{sh}}$ (see FIG.7)\par
\begin{eqnarray*}
\lefteqn{\sum_{j_i,m_i}(\prod_{i}w^2_{j_i})(-1)^{\sum_{i}m_i}
\sum_{J_A,m_A}(\prod_{A}w^2_{J_A}) (-1)^{\sum_{A}m_A}
\left( \begin{array}{ccc}
j_1 & j_2 & J_a \\
m_1 & m_2 & m_a
\end{array}\right)
\cdot}\hspace{.2cm}\nonumber \\
& & \cdot\left( \begin{array}{ccc}
j_3 & j_4 & J_a \\
m_3 & m_4 & -m_a
\end{array}\right)
\left( \begin{array}{ccc}
j_5 & j_6 & J_b \\
m_5 & m_6 & m_b
\end{array}\right)
\left( \begin{array}{ccc}
j_3 & j_7 & J_b \\
-m_3 & -m_7 & -m_b
\end{array}\right)\cdot\\
& & \cdot\left( \begin{array}{ccc}
j_5 & j_8 & J_c \\
-m_5 & m_8 & m_c
\end{array}\right)
\left( \begin{array}{ccc}
j_1 & j_9 & J_c \\
-m_1 & m_9 & -m_c
\end{array}\right)
[J_a,J_b,J_c,{\bf J_d},{\bf J_e}]=\\
& & =
w^2_L \sum_{m_{10}}(-1)^{m_{10}}
\sum_{m_B}(-1)^{\sum_{B}m_B}
\left( \begin{array}{ccc}
j_6 & j_{10} & J_d \\
-m_6 & m_{10} & m_d
\end{array}\right)
\left( \begin{array}{ccc}
j_2 & j_8 & J_d \\
-m_2 & -m_8 & -m_d
\end{array}\right)\cdot\\
& & \cdot \left( \begin{array}{ccc}
j_7 & j_{10} & J_e \\
-m_7 & -m_{10} & m_e
\end{array}\right)
\left( \begin{array}{ccc}
j_4 & j_9 & J_e \\
-m_4 & -m_9 & -m_e
\end{array}\right)
\end{eqnarray*}

$\{j_i\}=(j_1,j_3,j_5)$

$\{m_i\}=(m_1,m_2,m_5)$

$\{J_A\}=(J_a,J_b,J_c)$

$\{m_A\}=(m_a,m_b,m_c)$

$\{m_B\}=(m_d,m_e)$

$w^2_L=\Lambda(L)$.

\vskip 1cm

\noindent ${\bf [4 \rightarrow 1]^4_{sh}}$ (see FIG.8)\par
\begin{eqnarray*}
\lefteqn{\sum_{j_i,m_i}(\prod_{i}w^2_{j_i})(-1)^{\sum_{i}m_i}
\sum_{J_A,m_A}(\prod_{A}w^2_{J_A}) (-1)^{\sum_{A}m_A}
\left( \begin{array}{ccc}
j_1 & j_2 & J_a \\
m_1 & m_2 & m_a
\end{array}\right)
\cdot}\hspace{.2cm}\nonumber \\
& & \cdot \left( \begin{array}{ccc}
j_3 & j_4 & J_a \\
m_3 & m_4 & -m_a
\end{array}\right)
\left( \begin{array}{ccc}
j_5 & j_6 & J_b \\
m_5 & m_6 & m_b
\end{array}\right)
\left( \begin{array}{ccc}
j_3 & j_7 & J_b \\
-m_3 & m_7 & -m_b
\end{array}\right)\cdot\\
& & \cdot \left( \begin{array}{ccc}
j_5 & j_8 & J_c \\
-m_5 & m_8 & m_c
\end{array}\right)
\left( \begin{array}{ccc}
j_1 & j_9 & J_c \\
-m_1 & m_9 & -m_c
\end{array}\right)
\left( \begin{array}{ccc}
j_6 & j_{10} & J_d \\
-m_6 & m_{10} & m_d
\end{array}\right)\cdot\\
& & \cdot \left( \begin{array}{ccc}
j_2 & j_8 & J_d \\
-m_2 & -m_8 & -m_d
\end{array}\right)
[J_a,J_b,J_c,J_d,{\bf J_e}]=\\
& & = w^6_L\,\sum_{m_e}(-1)^{m_e}
\left( \begin{array}{ccc}
j_7 & j_{10} & J_e \\
-m_7 & -m_{10} & m_e
\end{array}\right)
\left( \begin{array}{ccc}
j_4 & j_9 & J_e \\
-m_4 & -m_9 & -m_e
\end{array}\right)
\end{eqnarray*}

$\{j_i\}=(j_1,j_2,j_3,j_5,j_6,j_8)$

$\{m_i\}=(m_1,m_2,m_3,m_5,m_6,m_8)$

$\{J_A\}=(J_a,J_b,J_c,J_d)$

$\{m_A\}=(m_a,m_b,m_c,m_d)$

$w^6_L=\Lambda(L)^3$.

\vskip 1cm
\noindent The full set of identities can be obtained, up to regularization, from 
anyone
of them, on applying orthogonality/completeness conditions for $3jm$ symbols 
as well as
orthogonality conditions for the $15j$ symbol (see \cite{lituani}).
Moreover, the set of inverse moves (corresponding 
to the attachment of a $4$-simplex) can be read in the same set of identities up 
to
exchanging the role of $J$ and ${\bf J}$.

\section*{Appendix B}

All state sums and associated classical invariants defined in terms of 
(re)coupling
coefficients of $SU(2)$ can be extended to the case of the quantum
enveloping algebra $U_q(sl(2,{\Bbb C}))$, $q$ a root of unity. 
Following the standard notation, the spin variables $\{j\}$ take their values in a 
finite set $I=(0,1/2,1,\ldots,k)$ where $exp(\pi i/k)=q$.
For each $j \in I$ a function $w^2_{(q)j} \doteq (-1)^{2x_j}[2x_j+1]_q \in K^*$ is 
defined, 
where $K^* \equiv K \setminus \{0\}$ ($K$ a commutative ring with unity). Recall 
that the notation $[.]_q$ stands for a $q$-integer, 
namely $[n]_q = (q^n-q^{-n})/(q- q^{-1})$
and that, for each admissible triple ($j,k,l$), we have: $w_{(q)j}^{-2} \sum_{k,l}
\,w^2_{(q)k} w^2_{(q)l} = w_q^2$, with $w_q^2=-2k/
(q-q^{-1})^2$.\par
\noindent  
We do not give the explicit expression of the quantum invariants  $Z[(M^d, \partial 
M^d)](q)$ and $Z[M^d](q)$ which would replace the classical counterparts found in 
{\em Section 5}
and {\em Section 6}. We just notice that the basic receipt to tranform the
classical state sums into the quantum ones can be summarized as follows\par

\begin{itemize}
\item    the classical 
weights $(-1)^{2j}(2j+1)$  are replaced by $w_{(q)j}^2$, while each of the factors 
$\Lambda(L)^{-1}$ becomes $w_q^{-2}$;

\item each Wigner symbol of $SU(2)$ is replaced by its $q$-analog $q-3jm$, 
normalized as explained below;   
\item each classical recoupling coefficient (or $3nj$ symbol) of a given type has 
its $q$-deformed counterpart, obtained by summing over magnetic numbers
products of $q-3jm$ symbols, apart from suitable phase factors.
\end{itemize}

Recall from \cite{kirillov} and \cite{nomura} that the relation between the 
quantum Clebsh--Gordan coefficient $(j_1m_1j_2m_2|j_3m_3)_q$
and the $q-3jm$ symbol is given by\par
\begin{equation}
(j_1m_1j_2m_2|j_3m_3)_q \,=\,(-1)^{j_1-j_2+m_3}
([2j_3+1]_q)^{1/2} \left(\begin{array}{ccc}
j_1 & j_2 & j_3 \\
m_1 & m_2 & -m_3
\end{array}\right)_q,
\end{equation}
\noindent where, as usual, an $m$ variable runs in integer steps between $-j$ and 
$+j$,
and the classical expression is recovered when $q=1$. 
The symmetry properties of the $q-3jm$ symbol read\par
\begin{eqnarray}
\left(\begin{array}{ccc}
j_1 & j_2 & j_3 \\
m_1 & m_2 & -m_3
\end{array}\right)_q = &
(-1)^{j_1+j_2+j_3} &
\left(\begin{array}{ccc}
j_2 & j_1 & j_3 \\
m_2 & m_1 & -m_3
\end{array}\right)_{1/q}, \nonumber\\
\left(\begin{array}{ccc}
j_1 & j_2 & j_3 \\
m_1 & m_2 & -m_3
\end{array}\right)_q = &
(-1)^{j_1+j_2+j_3} q^{-m_1/2} &
\left(\begin{array}{ccc}
j_1 & j_3 & j_2 \\
m_1 & m_3 & -m_2
\end{array}\right)_{1/q}, \nonumber\\
\left(\begin{array}{ccc}
j_1 & j_2 & j_3 \\
m_1 & m_2 & -m_3
\end{array}\right)_q = &
(-1)^{j_1+j_2+j_3} &
\left(\begin{array}{ccc}
j_1 & j_2 & j_3 \\
-m_1 & -m_2 & m_3
\end{array}\right)_q.
\end{eqnarray}
\noindent Thus we define the {\em normalized} $q-3jm$ symbols, 
for deformation parameters $q$ and $1/q$ respectively, according to\par
\begin{eqnarray}
\left[\begin{array}{ccc}
j_1 & j_2 & j_3 \\
m_1 & m_2 & -m_3
\end{array}\right]_q  &
\doteq q^{(m_1-m_2)/6}
\left(\begin{array}{ccc}
j_1 & j_2 & j_3 \\
m_1 & m_2 & -m_3
\end{array}\right)_q \nonumber\\
\left[\begin{array}{ccc}
j_1 & j_2 & j_3 \\
m_1 & m_2 & -m_3
\end{array}\right]_{1/q}  &
\doteq q^{(m_2-m_1)/6}
\left(\begin{array}{ccc}
j_1 & j_2 & j_3 \\
m_1 & m_2 & -m_3
\end{array}\right)_{1/q} 
\end{eqnarray}
\noindent The orthogonality relation involving the normalized symbols
(which are used for instance in order to handle identities representing elementay 
shellings
and inverse shellings) reads\par
\begin{equation}
\sum_{jm}w^2_{(q)j} (-1)^{\theta}q^{(m_2-m_1)/3}
\left[\begin{array}{ccc}
j_1 & j_2 & j \\
m_1 & m_2 & -m
\end{array}\right]_q
\left[\begin{array}{ccc}
j_2 & j_1 & j \\
-m'_2 & -m'_1 & -m
\end{array}\right]_q =
\delta_{m_1m'_1}\,\delta_{m_2m'_2},
\label{ortq3jm}
\end{equation}
\noindent where $\theta =m_1+m_2+m_3$.

\end{document}